\documentclass[fleqn,usenatbib]{mnras}
\usepackage[utf8]{inputenc}
\usepackage{amsfonts}
\usepackage{graphicx}
\usepackage{grffile}
\usepackage{amsmath}
\usepackage{amssymb}
\usepackage{newtxtext}

\usepackage{lmodern}
\usepackage{hyperref}
\usepackage{array}
\usepackage{listings}
\usepackage{comment}
\usepackage{bm}
\usepackage{slashed}
\usepackage{mathtools}

\usepackage{siunitx}
\usepackage{multirow}
\usepackage{booktabs}
\usepackage[para]{threeparttable}
\usepackage{ulem}
\usepackage{xcolor}
\newcommand{\refeqs}[2]{Eqs.~(\ref{eq:#1})--(\ref{eq:#2})}

\newcommand{\be}{\begin{equation}}
\newcommand{\ee}{\end{equation}}
\newcommand{\bea}{\begin{eqnarray}}
\newcommand{\eea}{\end{eqnarray}}
 
\def\ba#1\ea{\begin{align}#1\end{align}}

\renewcommand{\ln}{\operatorname{ln}}

\renewcommand{\P}{\mathcal{P}}

\def\tt{\tilde t_{0}}

\definecolor{RedWine}{rgb}{0.743,0,0}
\definecolor{GrassGreen}{rgb}{0.125,0.75,0.125}
\definecolor{RoyalBlue}{rgb}{0.25,0.41,0.88}
\definecolor{DarkCyan}{rgb}{0,0.5,0.5}

\newcommand{\cc}{{\cal C}}
\newcommand{\co}{{\rm col}}

\newcommand{\hto}{{\rm H_2}}

\newcommand{\hd}{{\rm HD}}

\numberwithin{equation}{section}

\renewcommand{\refeq}[1]{Eq.~(\ref{eq:#1})}
\renewcommand{\refeqs}[2]{Eqs.~(\ref{eq:#1})--(\ref{eq:#2})}
\title[Gravitational Collapse Induced by Cooling]{An Analytic Model of Gravitational Collapse Induced by Radiative Cooling: Instability Scale, Density Profile, and Mass Infall Rate}

\author[J. Gurian et al.]{James Gurian\textsuperscript{\href{https://orcid.org/0000-0002-8677-1038}{\includegraphics[width=2.5mm]{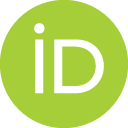}}\,}$^1$,\thanks{E-mail: jgurian@perimeterinstitute.ca}
Boyuan Liu\textsuperscript{\href{https://orcid.org/0000-0002-4966-7450}{\includegraphics[width=2.5mm]{orcid.png}}\,}$^{2,11}$,
Donghui Jeong\textsuperscript{\href{https://orcid.org/0000-0002-8434-979X}{\includegraphics[width=2.5mm]{orcid.png}}\,}$^{3,4}$,
Takashi Hosokawa\textsuperscript{\href{https://orcid.org/0000-0003-3127-5982}{\includegraphics[width=2.5mm]{orcid.png}}\,}$^5$,
Shingo Hirano\textsuperscript{\href{https://orcid.org/0000-0002-4317-767X}{\includegraphics[width=2.5mm]{orcid.png}}\,}$^{6,7}$,
\newauthor
Naoki Yoshida\textsuperscript{\href{https://orcid.org/0000-0001-7925-238X}{\includegraphics[width=2.5mm]{orcid.png}}\,}$^{8,9,10}$\\
$^1$ Perimeter Institute for Theoretical Physics, Waterloo, Ontario, N2L 2Y5, Canada\\
$^2$ Institute of Astronomy, University of Cambridge, Madingley Road, Cambridge, CB3 0HA, UK\\
$^3$ Department of Astronomy and Astrophysics and Institute for Gravitation and the Cosmos,\\ ~The Pennsylvania State University, University Park, PA, 16802, USA\\
$^4$School of Physics, Korea Institute for Advanced Study (KIAS), 85 Hoegiro, Dongdaemun-gu, Seoul, 02455, Republic of Korea\\
$^5$Department of Physics, Kyoto University, Sakyo, Kyoto 606-8502, Japan\\
$^6$Department of Astronomy, School of Science, University of Tokyo,  7-3-1 Hongo, Bunkyo, Tokyo 113-0033, Japan\\
$^7$Department of Applied Physics, Faculty of Engineering, Kanagawa University, Kanagawa 221-0802\\
$^8$Department of Physics, School of Science, The University of Tokyo, 7-3-1 Hongo, Bunkyo, Tokyo 113-0033, Japan\\
$^9$Research Center for the Early Universe, School of Science, The University of Tokyo, 7-3-1 Hongo, Bunkyo, Tokyo 113-0033, Japan\\
$^{10}$Kavli Institute for the Physics and Mathematics of the Universe (WPI), The University of Tokyo, Kashiwa, Chiba 277-8583, Japan\\
$^{11}$Institut für Theoretische Astrophysik, Zentrum für Astronomie, Universität Heidelberg, Albert Ueberle Straße 2, D-69120 Heidelberg, Germany
}
\date{Accepted XXX. Received YYY; in original form ZZZ}
\pubyear{2024}
\begin{document}

\label{firstpage}
\pagerange{\pageref{firstpage}--\pageref{lastpage}}
\maketitle

\begin{abstract}
We present an analytic description of the spherically symmetric gravitational collapse of radiatively cooling gas clouds, {which illustrates the mechanism by which radiative cooling induces gravitational instability at a characteristic mass scale determined by the microphysics of the gas}. The approach is based on developing the density-temperature relationship of the gas into a full dynamical model. We convert the density-temperature relationship into a barotropic equation of state, based on which we develop a refined instability criterion and calculate the density and velocity profiles of the gas. From these quantities we determine the time-dependent mass infall rate onto the center of the cloud. This approach distinguishes the rapid, quasi-equilibrium contraction of a cooling gas core to high central densities from the legitimate instability this contraction establishes in the envelope.  We explicate the model in the context of a primordial mini-halo cooled by molecular hydrogen, and then provide two further examples: a delayed collapse with hydrogen deuteride cooling and the collapse of an atomic cooling halo. In all three cases, {we show that} our results agree well with full hydrodynamical treatments. 
\end{abstract}


\section{Introduction}
Gravitational collapse leads to the formation of objects (e.g.~stars, degenerate stars, black holes, and planets) with densities tens of orders of magnitude above the cosmic mean. The physics relevant to the collapse include, at the bare minimum, gravitation, thermal pressure, and radiative cooling. Historically these dynamics could be modelled only under very restrictive assumptions amenable to analytic or\textemdash by modern standards\textemdash quite primitive numerical techniques. Today, the relevant physical processes can be included in great detail in sophisticated numerical simulations. Such studies have yielded powerful insights into the physics of gravitational collapse and star formation in a wide range of environments. However, the very complexity of these simulations can obscure the physical interpretation of the results. Moreover, there is increasing interest both in dark matter models which modify the gas collapse and star formation processes (for example due to exotic energy injection, \citealt{Ripamonti_2007,Freese_2016,qin2023birth}) and in the possibility that dark matter could itself cool and collapse to form dark compact objects \citep{D_Amico_2017,Shandera_2018,Chang_2019,Hippert_2022,Gurian:2022nbx,Bramante:2024pyc,Bramante:2023ddr}. In the face of large model and parameter spaces, state-of-the-art numerical treatments become rapidly intractable. It is thus a desirable goal to synthesize and distill the lessons learned from state-of-the art simulations into expository theories which (thanks to enormous advances in computational power) no longer need be restricted to such extremely idealized situations. {A particularly appealing theoretical target is the characteristic mass of gravitationally unstable clouds in which these objects form. While the mass function of the eventual collapsed objects depends on various complex physical processes, the typical mass of the natal collapsing clouds imposes an overall scale on the problem.}

Analytic descriptions and heuristics describing the collapse of gases governed by simple equations of state (i.e.~isothermal or polytropic) are well established in the literature (e.g.~\citet{Larson69, Penston1969,Hunter77, Shu77}). We mention in particular two similarity solutions: the first derived by \citet{ Larson69} and \citet{Penston1969} (hereafter referred to as the Larson-Penston solution) and the second by \citet{Shu77} (the Shu solution). The former describes the highly dynamical collapse  of a Bonnor-Ebert sphere, while the latter describes the quasi-static collapse of a singular isothermal sphere triggered by the propagation of a rarefaction wave after core formation. Simulations typically reveal an intermediate picture, where the gas is accelerated towards the Larson-Penston solution over the  course of the collapse \citep{Foster93, McKee2002,McKee2003,Tan2004,Omukai2010}. 

While similarity solutions are exact under the appropriate assumptions, they are by definition scale-free. That is to say they provide no information about the beginning, end, {or mass scale} of the collapse. The canonical scale associated with the onset of gravitational collapse is the Jeans scale, which describes a balance between pressure gradients and gravity \citep{Jeans28}, given as 
\begin{equation}
    M_J \approx 1.44\left(\frac{ k_B  \bar T}{\mu m_P G }\right)^{3/2}\bar\rho^{-1/2},
\end{equation}  
where $k_B$ is the Boltzmann constant, $\bar T$ the average temperature, $\bar \rho$ the average density, $\mu$ the mean molecular weight, and $G$ the gravitational constant. The intuition is that at small scales pressure damps out perturbations while at large scales gravity overwhelms pressure support. This argument was later refined by \cite{Bonnor1956,Ebert55} as the Bonnor-Ebert mass, discussed in detail below. Calculating the Jeans mass requires a fixed density and temperature. To use the Jeans mass to pick out a scale for the onset of gravitational collapse is justified when the spatial density structure of the gas is independent of its pressure, for example if the density probability distribution is set by the statistics of turbulence~\citep[i.e.][]{Hopkins_2012,Hopkins_2013}. 

In fact, as the density in a gas cloud increases the Jeans mass will decreases as long as $T$ increases more slowly than $\rho^3$. A consequence is that over the course of the collapse, progressively smaller scales can become unstable. This process of ``hierarchical fragmentation'' is ultimately terminated when the gas becomes optically thick and unable to cool efficiently. For stars, the opacity limit is order $10^{-3} \, \rm M_\odot$ \citep{Rees76}. Still, the opacity-limited Jeans mass has often been adopted in the dissipative dark matter literature as a heuristic for the final mass of the hydrostatic objects produced by the collapse, either directly  \citep{Chang_2019,Bramante:2023ddr, Bramante:2024pyc,Fernandez_2024}, as a lower bound \citep{Gurian:2022nbx}, or with a constant multiplicative enhancement \citep{Shandera_2018}. 

An alternative argument picks out a preferred scale in the collapse based on deviations from isothermality, which alter the effective equation of state of the gas. It is widely appreciated that when the temperature is an increasing function of density fragmentation is suppressed, while when temperature is a decreasing function of density fragmentation is enhanced \citep{Larson1985,Larson2005,Li_2003}. These observations are theoretically best justified in the case of filamentary geometries \citep{Ostriker64,Inutsuka92,Omukai2005}. On the other hand, fragmentation in the sense of growth of initially small perturbations at some preferred scale has been shown to be ineffective during global free-fall collapse \citep{Bodenheimer1980,Tohline1980,Tohline1980b}. 

Neither Jeans-based argument explicitly considers the (in)efficiency of radiative cooling. Without cooling, a Jeans unstable cloud compressionally heats to a new equilibrium. On the other hand, in the presence of efficient radiative cooling even a Jeans stable cloud will contract on its cooling timescale, which may be comparable to its free-fall timescale \citep{bromm_forming_1999,gurian2024zero}. Radiative cooling is explicitly accounted for in the Rees-Ostriker criterion \citep{Rees77}. The argument is that a gas which can cool within its dynamical (free-fall) timescale will undergo dynamical collapse and fragmentation. This calculation requires single, characteristic values for the temperature and density. In general, the gas will have some density, temperature, and chemical composition gradients. The cooling and free-fall timescales can be quite sensitive, non-linear functions of these quantities. It is not obvious that a naive average of these quantities over some region will produce a physically reasonable mass scale for the onset of the collapse. \citet{Bertschinger1989} and \citet{White1991} accounted for this fact by calculating the ``cooling radius,'' defined as the radius at which the local cooling time (in some assumed density profile) equals the age of the system. In particular, \citet{Bertschinger1989} discovered a similarity solution based on this length scale for the evolution of the cooling gas. However, the similarity exists only for power-law (i.e.~scale-free) density and pressure profiles. Cooling can modify the effective equation of state of the gas in a scale dependent manner, which limits the applicability of the solution. {Moreover, the scale at which cooling becomes efficient does not necessarily correspond to the onset of gravitational instability or fragmentation. An efficiently cooling core of a gas cloud can remain quasi-hydrostatic in structure if the sound crossing time is sufficiently short, while fragmentation depends on small-scale density perturbations.}

{As detailed above, this extensive prior work does not fully succeed in defining the characteristic mass scale of gravitational instability in all contexts. The shortcomings of these heuristics are clearly illustrated in the context of the formation of first generation, Pop.~III stars. In pristine (metal-free) gas, the only significant coolants are molecular hydrogen ($\rm H_2$), hydrogen deuteride ($\hd$), and atomic hydrogen ($\rm H$) \citep{Liu2018}. In the canonical case of mini-haloes cooled by molecular hydrogen, gravitational instability has long been associated with the Jeans scale at the minimum temperature over the course of the collapse, i.e., the ``loitering point'' \citep{bromm_forming_1999}, $\sim 10^{3} \, \rm M_\odot$. This minimum temperature occurs at the critical density of molecular hydrogen, where collisional de-excitation begins to compete with radiative de-excitation. However, radiative cooling typically becomes efficient (in the sense that the cooling timescale becomes as short as the dynamical timescale) at a lower density and larger mass scale $10^4 $--$10^5\, \rm M_\odot$. Moreover, the early phase of the collapse is monolithic, with typically only one star forming cloud per halo. That is, fragmentation into multiple Jeans-scale clumps does not actually occur and should not be invoked as an explanation for this characteristic mass. }

{Here we show that the characteristic scale of this gravitational instability can be explained by the non-homologous nature of the collapse, as a rapidly cooling (Rees-Ostriker unstable) but perturbatively (Jeans/Bonnor-Ebert) stable core of gas contracts and establishes an out-of-equilibrium density profile in its envelope. The degree of instability in this envelope determines the subsequent infall rate onto the proto-star and its disc. To this end, we develop a dynamical model of gravitational collapse which explicitly includes thermal pressure, gravity, and radiative cooling. As our test-bed, we consider the collapse of primordial gas into first generation (Pop.~III) stars, where the initial conditions for the collapse are dictated by the cosmological environment, and can be described in terms of a relatively small number of physical quantities. Still, a wide range of outcomes are possible for the collapse, and the resulting Pop.~III initial mass function (IMF) remains a topic of active research (for reviews see \citealt{Bromm2004, Bromm:2013, Haemmerl__2020, klessen2023stars}). As discussed above, the radiative cooling physics are thought to play a crucial role in setting the mass-scale of the gravitational collapse. In addition to molecular hydrogen, the formation of deuterated hydrogen (which has a permanent dipole moment) leads to a lower minimum temperature and less massive collapsing cloud \citep{Ripamonti_2007, hirano_one_2014,nishijima2023lowmasspopiiistar}, while nearly isothermal atomic cooling is associated with direct collapse and the formation of supermassive stars \citep{Omukai2005,Latif_2013, Wise_2019,Kiyuna_2023}. }

{In all these cases, cooling remains efficient until the formation of a protostar. For this reason, the density in the centre of the cloud rapidly increases \textit{independent} of the gravitational stability of the cloud. As the density increases, the core becomes both smaller and less massive. If cooling remains efficient indefinitely, the endpoint of the contraction phase is an infinitely concentrated and infinitesimally small core. It is this core-contraction which can (but does not necessarily) establish true gravitational instability in the surrounding envelope. We demonstrate here that the mass scale of this gravitational instability in the envelope is controlled by features in the temperature-density relationship.}

{Our model uses the density and temperature dependent radiative cooling rates to determine the quantity of gravitationally unstable gas ``left behind'' by the core-contraction, and to estimate the rate at which this gas will fall onto the proto-star, or its accretion disc.} The model is based on defining an effective barotropic equation of state for the gas from the thermal evolution in the core, {which in this work we supply using a one-zone model}. We demonstrate the importance of a modified Bonnor-Ebert scale in regulating the contraction of the core. We use this scale to calculate a radial density profile for the gas, including both the pressure supported core and the envelope established by the core contraction. We assess the gravitational (in)stability of the envelope by the ratio of the mass enclosed to the modified Bonnor-Ebert mass, $\kappa_{\rm MBE}$. Finally, we determine the time-dependent mass infall rate from the envelope onto the central hydrostatic core, which we connect to $\kappa_{ \rm MBE}$. The model preserves the physical transparency and computational expediency of analytic approaches while including the full temperature and density dependence of the relevant cooling rates. 

The calculation is perhaps most similar in spirit to the series of papers \citet{Sipil__2011, Sipil__2015, Sipil__2017}, which calculated a Bonnor-Ebert stability criterion for pre-stellar gas clouds using numerically determined density and temperature profiles for the clouds. Where those works determined the ``critical'' (marginally stable) central density of gas cores of fixed mass, we undertake a dynamical model of the collapse based on a sequence of marginally stable cores. The details of the implementations also differ: where those works determined the density and temperature profiles using an iterative procedure involving 1D radiative transfer, we employ an effective barotropic equation of state generated by a one-zone calculation. Then, we determine the density and temperature profile by numerically solving a sequence of ordinary differential equations.

We also mention the recent work of \citet{smith2024does}, which illustrates the importance of radiative cooling in controlling the onset of gravitational instability by demonstrating a critical gas-phase metallicity for star formation in strong UV backgrounds. That work uses a combination of three-dimensional simulations and one-zone modelling. In the one-zone model, the density and temperature can be understood as average values. Gravitational collapse (and thus star formation) is assessed to begin when the one-zone density and temperature indicate instability via the isothermal Bonnor-Ebert criterion. Here, we develop a modified Bonnor-Ebert condition, which can account for  temperature/pressure gradients as well as the contribution of dark matter to the gravitational potential. Based on this Bonnor-Ebert criterion, we build up a one-dimensional model of the dynamics of the collapse and show that the onset of instability can be understood through the thermal evolution of the gas. We explain the qualitative agreement between our model (in which the gas core is never unstable) and mean density based calculations such as that of \citet{smith2024does} in Appendix \ref{app:mjratio}.

Our approach is tractable in large part due to the powerful tools provided by the SciML ecosystem \citep{rackauckas2017differentialequations} for solving and analyzing differential equations.

This paper is organized as follows. In Section \ref{section:GC}, we develop the model using the canonical example of a mini-halo cooled by molecular hydrogen. Subsequently, we apply these methods to two further examples in Section \ref{sec:ex}. In Section \ref{sec:delay} we consider the case where a delayed collapse leads to the formation of $\hd$ which delays gravitational instability to higher density and smaller mass. In Section \ref{sec:atomic} we consider the opposite case, where the gas heats up to the point that atomic cooling is efficient. There, the nearly-isothermal equation of state leads to prompt gravitational collapse. We close with a summary of the main results and brief discussion of directions for future research.

\section{Method}
\label{section:GC}
We begin by explicating the model using the example of a mini-halo cooled by molecular hydrogen, before turning to further examples in the next section. The steps of the calculation are as follows. We first generate an effective barotropic equation of state for the gas using a one-zone calculation (Section \ref{sec:eos}), and then apply this equation of state to compute a radial density profile valid in the inner, pressure-supported part of the cloud, which we first use to generalize the Bonnor-Ebert stability condition and apply this condition to calculate the full density profile of the gas (Section \ref{sec:profilebe}). We discuss the gravitational stability of this density profile by calculating the ratio of the mass enclosed to the Bonnor-Ebert mass and, in Section \ref{sec:infall}, by calculating the time-dependent mass accretion from the envelope onto the core.

\subsection{The Effective Equation of State}
\label{sec:eos}
{Our model requires an effective equation of state as an input. Here, we calculate this equation of state using the density-temperature relationship determined from a one-zone model. This equation of state could also in principle be determined by timescale arguments (e.g.~\citealt{Chang_2019,Bramante:2023ddr}), or from the density temperature-relationship in marginally resolved structures in a hydrodynamical simulation. The underlying logic is that due to the self-regulatory behavior of the thermal evolution (discussed below) even a simple ansatz for the density evolution can produce a reasonably accurate density-temperature relationship and hence effective equation of state, which we will here build into a more accurate dynamical model. } We begin by briefly explicating the one-zone model, and refer the reader to e.g.~\citet{gurian2024zero} for more detailed discussion. The temperature evolution of a uniform density parcel of gas (say, in the core of a gas cloud) is given as 
\begin{equation}
      \frac{dT}{dt} = (\gamma -1)\left[\frac{ \dot n}{n}T-\frac{{\cal C}(T,{\vec n})}{k_B n} \right],
    \label{eq:Tevol}
\end{equation}
with $\gamma$ the adiabatic index,  $n$ the total number density, $T$ the temperature, $\cal C$ the volumetric cooling rate and $\vec{n}$ the number densities of the various species. Evaluating this equation at a given density requires the chemical composition $\vec{n}$ and the time derivative of the density. The former can be supplied by solving a chemical network (i.e.~a system of ordinary differential equations describing the interconversion of the various species). However, calculating $\dot n$ requires the full dynamics of the gas, including gravitation and pressure. These dynamics can (at considerable computational cost) be supplied by hydrodynamical simulations. Instead, we apply a simple ansatz the the density in our gas parcel evolves on some characteristic collapse timescale: 
\begin{equation}
\dot \rho = \frac{\rho}{t_{\co}}.
\label{eq:drho}
\end{equation}

Under this assumption, we can numerically integrate \refeq{Tevol} together with the chemical network and determine $T$ as a function of $\rho$ alone. Note that \refeq{Tevol} can be rewritten as \citep{gurian2024zero}
\begin{equation}
    \frac{d \log T}{d \log n}= (\gamma -1)\left[
    1 -\frac{t_{\co}(n)}{(\gamma-1)t_{\cc}({\vec n},T)}\right],
    \label{eq:evol}
\end{equation}
which demonstrates a self-regulatory behavior of the gas, in that over the course of the collapse the temperature will adjust so that $t_{\co}(n) \approx (\gamma -1) t_{\cc}$. 

\begin{figure}
    \centering
    \includegraphics[width=\columnwidth]{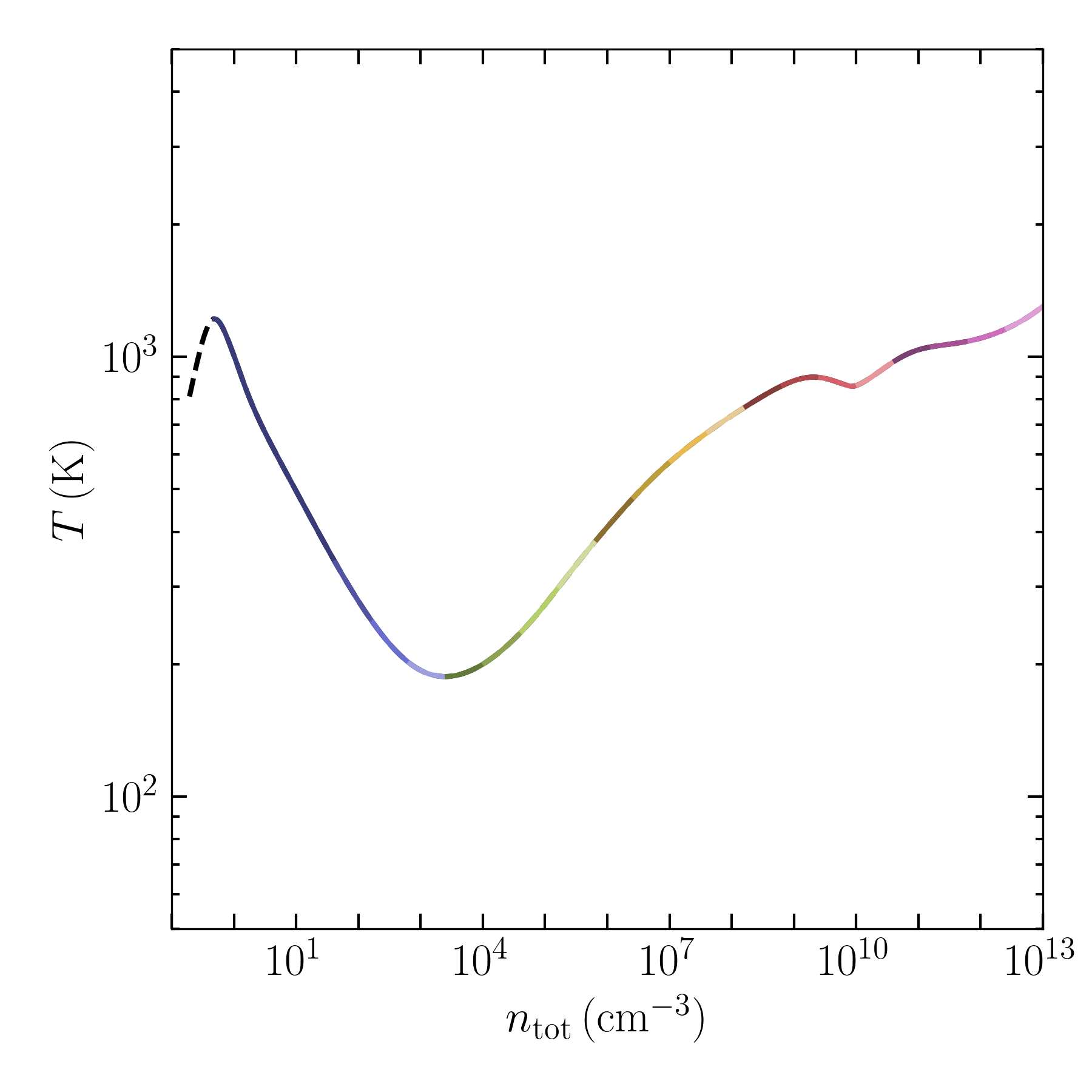}
    \vspace{-20pt}
    \caption{The one-zone temperature density relationship for a $5 \times 10^5 \, \rm M_\odot$ halo at $z=25$ collapsing on the free-fall timescale. The black, dashed part of the curve is the initial, adiabatic heating up to the point where $t_{\co} \sim t_{\cal C}$. The cooling part of the curve is color-coded for easy identification with subsequent figures. }
    \label{fig:phaseh2}
\end{figure}

For the example molecular cooling mini-halo, we adopt the initial abundances described in Tab.~\ref{tab:abundances} and solve a standard chemical network using \textsc{krome} \citep{grassi_krome_2014} with the  initial temperature and density set appropriate to a $5 \times 10^5 \, \rm M_\odot$ halo at $z=25$, taking $t_{\co} = t_{\rm ff}$, with the free-fall timescale defined by 
\begin{equation}
    t_{\rm ff} = \sqrt{\frac{3 \pi}{32 G \rho}}.
\end{equation}
{Throughout this work we will take $t_{\rm col} = t_{\rm ff}$ (in this example and in Section \ref{sec:atomic}) or $t_{\rm col} = f t_{\rm ff}$ (in Section \ref{sec:delay}), with $f$ a constant, here effectively a free parameter. We discuss this choice and the possibility of time-varying $f$ in App.~\ref{app:fvar} }
The resulting density-temperature relationship is shown in Fig.~\ref{fig:phaseh2}. At low densities and temperatures, the cooling is not yet efficient and the gas evolves by adiabatic heating. The first local maximum of the temperature is the intersection of the thermal trajectory of the gas with the curve $t_{\co}(n) \approx (\gamma -1) t_{\cc}$. This marks the beginning of the cooling-regulated core contraction, {and is closely related to the Rees-Ostriker condition. The difference is that we allow $t_{\co}$ to vary by an overall, order unity factor from $t_{\rm ff}$, which is an approximate treatment of any slowdown in the evolution due, for example, to temporarily ineffective cooling or rotational support. This runaway contraction on a timescale comparable to $t_{\rm ff}$ continues until thermal pressure overcomes the gravitational force, when the adiabatic index $d \ln p/d\ln\rho >4/3$ (e.g. \citealt{Omukai2005}). } In the remainder of this work we do not consider the initial, heating part of the trajectory. We integrate the chemical-thermal network until the central density reaches $10^{13} \, \rm cm^{-3}$, a number chosen somewhat arbitrarily but far larger than the ``loitering point'' which is our primary interest. 
\begin{table}
    \centering
    \caption{The initial fractional abundances and their sources. \label{tab:abundances}}
    \begin{tabular}{cll}
    \hline\hline
        Species & Initial Abundance & Source\\
        \hline
        $x_e$ & $2.5\times 10^{-4}$& {\textsc{recfast}} \citep{Seager_1999}\\
        $x_{\hto}$ & $7 \times 10^{-7}$&\citet{Hirata2006}\\ 
        $x_{\rm D}$ &$2.5 \times 10^{-5}$&\citet{Cooke2018}\\
        $x_{\rm D^+}$ & $6.3 \times 10^{-9}$& $x_{\rm D+}/x_{\rm D} \equiv x_{\rm H+}/x_{\rm H}$\\
        $x_{\rm HD}$ & $1.8\times 10^{-11}$&$x_{\rm HD}/x_{\rm D} \equiv x_{\rm H_2}/x_{\rm H}$\\
        \hline\hline
    \end{tabular}
\end{table}
\subsection{Density Profile and Bonnor-Ebert Mass}
\label{sec:profilebe}
Based on this density-temperature relationship, we wish to calculate a radial density profile. The guiding intuition is that for any given central density and temperature, the radius of the gas core is of order the local Jeans length. Further, the collapse is highly non-homologous in the sense that the density far from the core hardly changes as the central density increases. Therefore, we can sketch a density profile by conceptually inverting
\begin{equation}
    r=\lambda_J(\rho, T(\rho)),
\end{equation}
with $\lambda_J$ the Jeans length. In this section, we develop this intuition using a modified Bonnor-Ebert scale.

\subsubsection{The Core Profile}
In the inner ``core'' region we determine the density profile by numerically integrating the equation of hydrostatic equilibrium. This is reasonable because even in the presence of efficient radiative cooling, pressure can regulate the collapse on small scales. In other words, sufficiently deep in the gas core, the sound crossing time is short compared to the evolutionary timescale. We will shortly determine the threshold where the quasi-hydrostatic evolution breaks down, which is the Bonnor-Ebert scale. Now, in spherical symmetry, the equation of hydrostatic equilibrium is
\begin{equation}
-\frac{G \left[M_{\rm DM}(r) + M(r)\right]\rho}{r^2} = \frac{dP}{d\rho}\frac{d\rho}{dr},    
\label{eq:hse}
\end{equation}
where $M_{\rm DM}$ is the dark matter mass, $M$ the gas mass, $\rho$ the gas density, and the pressure $P$ and its derivative are supplied by the effective barotropic equation of state
\begin{equation}
    P(n) =  n k_B T(n).
\end{equation}
\refeq{hse} can be integrated numerically along with the equation of mass conservation 
\begin{equation}
    \frac{d M_{\rm tot}}{dr} = 4\pi r^2 \left[\rho(r)+\rho_{DM}(r)\right],
    \label{eq:mcons}
\end{equation}
where $M_{\rm tot} =M_{\rm DM} + M$, and we are neglecting the effect of the gas evolution on the dark matter (though see \citealt{Spolyar2008}). In our molecular cooling mini-halo, we take a dark matter density profile informed by the simulations of \citet{hirano_one_2014}, which generated a sample of $\sim 100$ clouds collapsing in haloes of masses between $10^5 \, \rm M_\odot$ and $10^6 \, \rm M_\odot$ and at redshifts between $10 \lesssim z \lesssim 35$. The dark matter density profiles found in that work can be approximated by \citep{hirano_talk}, 
\begin{equation}
    \rho_{\rm DM}(r) = \frac{\sqrt 2 \rho_s}{\left(\frac{r}{r_s}\right)^{3/2}\left(1+ \frac{r}{r_s}\right)^{1/2}},
    \label{eq:dmsim}
\end{equation}
with $\rho_s=5 m_{\rm H} \,\rm cm^{-3},$ where $m_{\rm H}$ is the mass of the hydrogen atom and $r_s= 30$ parsecs. The parameters $\rho_s$ and $r_s$ should in principle depend on the halo mass and redshift, and at a given radius the density can vary by a factor of $\sim 10$ over the simulation suite. We discuss the sensitivity of our results to the dark matter density in Appendix \ref{app:dmprofile}.

\begin{figure}
    \centering
    \includegraphics[width=\columnwidth]{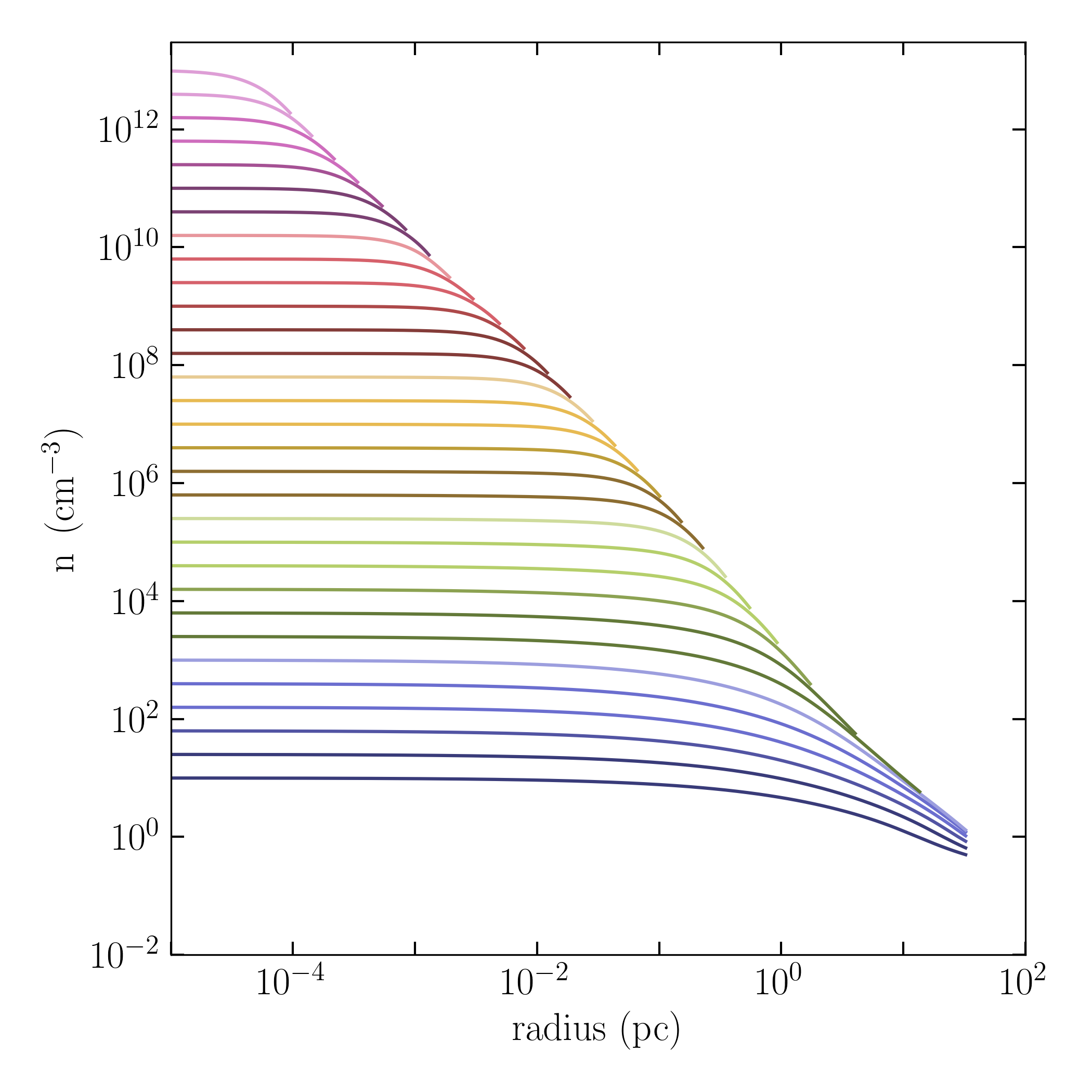}
    \vspace{-20pt}
    \caption{The hydrostatic density profiles out to the mininum of the Bonnor-Ebert radius or the initial values of $n$, $T$ in Fig.~\ref{fig:phaseh2}, for a range of central densities. The central density for each curve is color-coded to correspond with Fig.~\ref{fig:phaseh2}. }
    \label{fig:h2coreprof}
\end{figure}

With $\rho_{\rm DM}$ set, we can numerically integrate \refeqs{hse}{mcons} from any assumed central density $\rho_c$. The result is the hydrostatic radial density profile of the gas, $\rho_{\rm HSE}(\rho_c,r)$. At large radii, the hydrostatic density profile can represent an unstable equilibrium. The transition from stable to unstable equilibrium is defined by the Bonnor-Ebert scale. In Fig.~\ref{fig:h2coreprof}, we plot the hydrostatic density profile out to whichever is less of the modified Bonnor-Ebert radius (defined below) or the initial temperature maximum of the density-temperature relationship (where $t_{\co} \sim t_{\cc}$, which also defines a cooling radius for the gas $r_{\cal C}$\footnote{We point out that this $r_{\cal C}$ differs conceptually from the cooling radius of e.g.~\citet{Bertschinger1989}, which is based on the age of the system rather than its dynamical timescale.}). 

\subsubsection{The Bonnor-Ebert Mass}
In the standard treatment of gravitational instability (in which radiative losses are neglected) excessively centrally concentrated density profiles are unstable to collapse and fragmentation while less concentrated profiles can remain indefinitely hydrostatic. The picture is different for gases which lose kinetic energy through radiation. Such clouds can quasi-hydrostatically contract to higher central densities regardless of the presence or absence of a perturbative instability. In this case, both the cooling (or Kelvin-Helmholtz) and free-fall timescales are decreasing functions of density. This results in a tendency for the dense, inner regions of a cloud to ``run away'' to still higher densities on the relevant local timescale. 

For such a collapsing central region in a (quasi-)static external medium, a Bonnor-Ebert stability criterion applies. Consider a central region of the cloud infinitesimally contracting to a new hydrostatic configuration on its local Kelvin-Helmholtz timescale. If this contraction leads to an increase in surface pressure, sound waves will push the gas back towards its original configuration. On the other hand, if the surface pressure decreases then the gas will be accelerated towards the center of the cloud. 

Exact contraction on the local Kelvin-Helmholtz timescale corresponds to marginal Bonnor-Ebert stability,
\begin{equation}
    \left(\frac{\delta P}{\delta V}\right)_M = 0,  
    \label{eq:BEC}
\end{equation}
which is the classical Bonnor-Ebert condition that applies to spherical gas clouds of fixed mass \citep{Bonnor1956,Ebert55}. However, the Bonnor-Ebert mass (like the Jeans mass) is usually a decreasing function of central density. This means that in the case of the contracting gas core, we should no longer hold the mass fixed. Thus, we consider instead the surface pressure response to an increase in central density, at fixed radius:
\begin{equation}
    \frac{\delta P}{\delta \rho_c} = 
    \left.\frac{\partial P}{\partial\rho}\right|_{\rho=\rho_{\rm HSE}}
    \frac{\partial \rho_{\rm HSE}}{\partial \rho_c}=0,
    \label{eq:dpdrhoc}\,
\end{equation}
which defines the radius/density at which contraction proceeds in pressure equilibrium. For the HSE case that we show in Fig.~\ref{fig:h2coreprof}, $\rho_c$ is one-to-one to the size of the core. As long as $\frac{\partial P}{\partial\rho}$ does not change sign\footnote{In fact, as $\frac{\partial P}{\partial \rho} \rightarrow 0$, the sound speed also goes to zero, promoting the development of shocks and a breakdown of our treatment. In all the cases considered here $\frac{\partial P}{\partial \rho} > 0$.}, the first zero of this quantity occurs when
\begin{equation}
    \frac{\partial\rho_{\rm HSE}}{\partial\rho_c} = 0.
    \label{eq:rbe}
\end{equation}
\refeq{rbe} defines the {modified Bonnor-Ebert (MBE) radius throughout the rest of this work.} In Appendix \ref{app:be}, we explicitly relate this condition to the standard Bonnor-Ebert condition, \refeq{BEC}. Note that \refeq{rbe} in fact requires calculating the derivative of a numerical solution to an ordinary differential equation with respect to its initial condition. This is readily accomplished using the sensitivity analysis tools provided as part of the SciML ecosystem \citep{rackauckas2020universal}. 

 As long as the cooling timescale is a decreasing function of density, during the contraction the outer regions of the cloud can never ``catch up'' to the contracting core. That is, the evolution of the inner part of the cloud is approximately independent of the outer part. The modified Bonnor-Ebert condition \refeq{rbe} sets the scale at which the central region of the gas cloud decouples from its surroundings and escapes to high densities via radiative cooling.

Associated with the modified Bonnor-Ebert radius $r_{\rm MBE}$ is the Bonnor-Ebert mass. Defining $\rho_{c,{\rm MBE}}\equiv \rho_{c}(r_{\rm MBE})$ as the central density for which the Bonnor-Ebert radius is $r_{\rm MBE}$,
\begin{equation}
    M_{\rm MBE} = 4\pi \int_0^{r_{\rm MBE}} r^2dr\, \rho_{\rm HSE}(\rho_{c,{\rm MBE}},r),
\end{equation}
which is the largest possible stable, hydrostatic mass enclosed in the Bonnor-Ebert radius $r_{\rm MBE}$, consistent with the effective barotropic equation of state of the gas. 

\subsubsection{The Envelope and Cloud Mass}
At a given central density $\rho_c$ and for $r \le r_{\rm MBE}(\rho_c)$ we calculate the density profile by solving the equation of hydrostatic equilibrium, \refeq{hse}: 
\begin{equation}
    \rho(r) = \rho_{\rm HSE}(\rho_c, r).
\end{equation}

For $r > r_{\rm MBE}(\rho_c)$, we calculate the density as the hydrostatic equilibrium density at that radius when $r_{\rm MBE}=r$ (that is, at some previous time when $\rho_c=\rho_{c,\rm MBE}$):
\begin{equation}
    \rho(r)=\rho_{\rm MBE}(r) \equiv \rho_{\rm HSE}(\rho_{c,\rm MBE},r).
\end{equation}
In other words, we assume that the density at $r(> r_{\rm MBE})$  has not evolved since the earlier time/lower central density where $r$ was the Bonnor-Ebert radius. The assumption is reasonable because when $\rho_c \approx \rho_{c,\rm MBE}$ the density at $r$ must be slowly evolving by \refeq{rbe}, while once the central density has increased such that $\rho_c \gg \rho_{c,\rm MBE}$ the central evolutionary timescale is very short compared to the evolutionary timescale at $r_{\rm MBE}$. Then, the full density profile is
\begin{equation}
    \rho(\rho_c, r) =  \begin{cases} 
      \rho_{\rm HSE}(\rho_c, r)& r\leq r_{\rm MBE}(\rho_c) \\
     \rho_{\rm MBE}(r)  & r > r_{\rm MBE}(\rho_c).
   \end{cases}
   \label{eq:envprof}
\end{equation}

\begin{figure}
\centering
\includegraphics[width=\columnwidth]{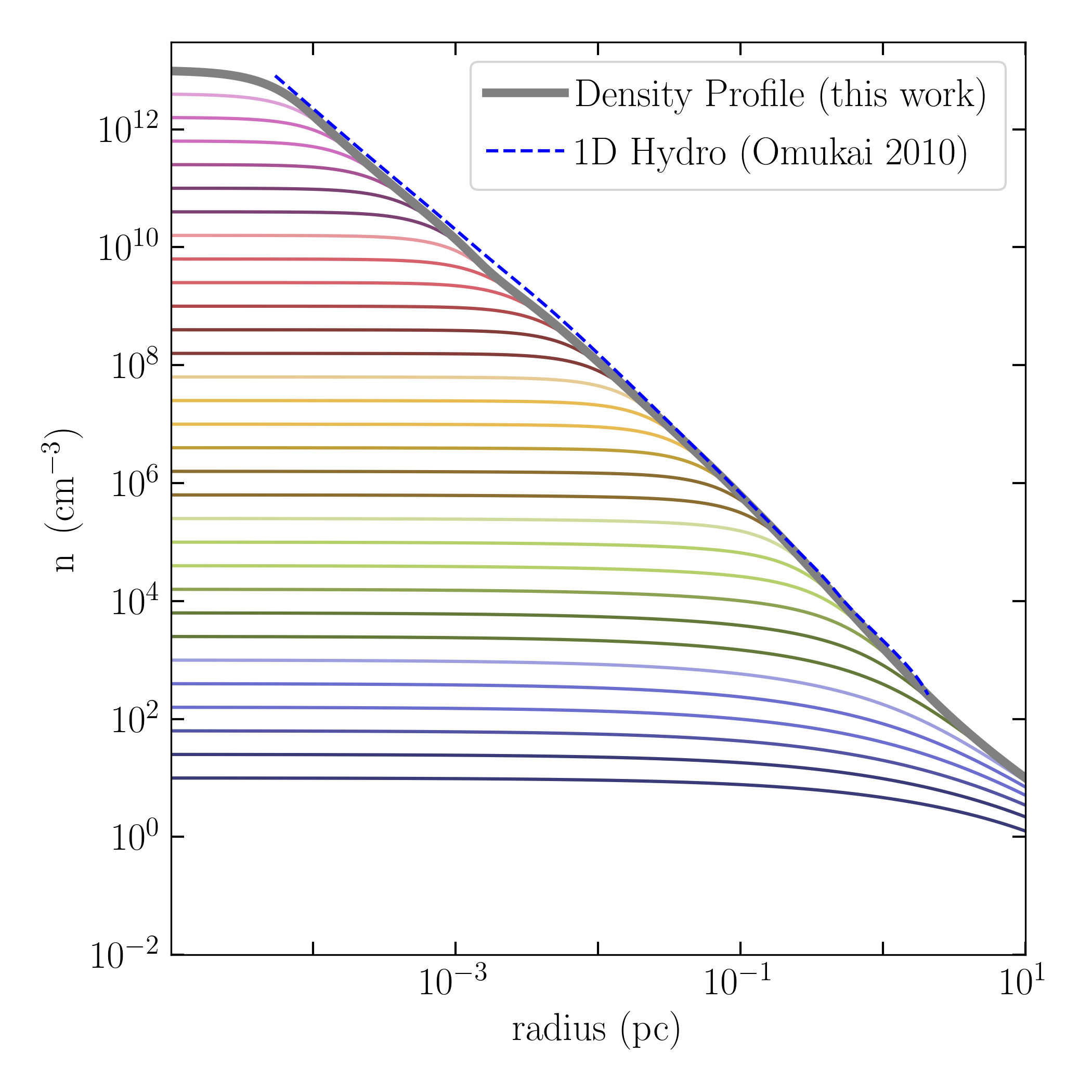}
\vspace{-20pt}
\caption{As Fig.~\ref{fig:h2coreprof}, but we have now overplotted in grey the full density profile (\refeq{envprof}) when the central density is $10^{13} \, {\rm cm}^{-3}$, as well as the late-time density profile from the 1D simulations of \protect\citet{Omukai2010} , here truncated at $10^{13} \, {\rm cm}^{-3}$. The agreement is quite close!}
\label{fig:h2fullprofile}
\end{figure}

In practice, we determine the density profile for $r>r_{\rm MBE}$ by calculating the hydrostatic density profiles and Bonnor-Ebert radii over a grid of central densities, then splining through the hydrostatic density profiles at the Bonnor-Ebert radius of each. This is illustrated in Fig.~\ref{fig:h2fullprofile}. In the inner region, the profile agrees closely with that derived from the 1D hydrodynamic calculations of \citet{Omukai2010}. {Although that work did not include dark matter, we show in Appendix \ref{app:dmprofile} that the effect of the dark matter on the density profile is minor in this inner region. Further, in Appendix \ref{app:sims} we compare the model with 3D simulations.} The slope is also consistent with the Larson-Penston polytropic solution for a gas with adiabatic index $\gamma \approx  1.1$ (which here holds between $10^4 {\, \rm cm^{-3}} \lesssim n\lesssim 10^7{\, \rm cm^{-3}}$) \citep{Omukai_1998}. Below this density, the  Larson-Penston solution predicts a shallower density profile, which is not seen here due to the dark matter dominating the density at large radii (see Appendix \ref{app:dmprofile}). 

Given these density profiles, for any central density and radius we can calculate the ratio of the mass enclosed to the Bonnor-Ebert mass, 
\begin{equation}
    \kappa_{\rm MBE} = \frac{M(r)}{M_{\rm MBE}(r)},
\end{equation}
for any $r$ inside of the cooling radius ($r \leq r_{\cal C}$). 

When $\kappa_{\rm MBE}<1$ for all $r \leq r_{\cal C}$, the gas evolves on a global timescale. If the low density gas (i.e.~the gas beyond $r_{\cal C}$) is adiabatically falling onto the core, this global timescale may still be comparable to the free-fall timescale. On the other hand, if the surrounding medium is nearly hydrostatic the timescale may be considerably longer. 

As soon as $\kappa_{\rm MBE}$ reaches unity {within $r_{\cc}$, the Bonnor-Ebert core begins to contract on its cooling timescale, with $t_{\cc} \sim t_{\rm ff}$. That is, a refined Rees-Ostriker criterion is given by:
\begin{equation}
    \kappa_{\rm MBE}(r_{\cc}) = 1.
\end{equation}} 

As the central density increases and the Bonnor-Ebert radius decreases the cooling and free-fall timescales become shorter still: a period of runaway Kelvin-Helmholtz contraction begins. {Throughout this phase, the density profile in the contracting core remains stable and nearly hydrostatic ($\kappa_{\rm MBE} \leq 1$).} The contraction begins to decelerate once the gas can no longer radiate its gravitational energy within a free-fall time {such that the equation of state becomes stiff with $d \log P/d\log  \rho >4/3$}, for example after becoming optically thick \citep{Rees76,Low76}. By this point, the contraction of the core, which has proceeded nearly in equilibrium, has established a new density profile in the envelope. As we will establish quantitatively in the following section (\ref{sec:infall}), the amount by which $\kappa_{\rm MBE}$ exceeds unity at a given radius is related to the subsequent infall rate {from the envelope onto the core}: a larger value indicates more violent acceleration towards the core.\footnote{In fact, $\kappa_{\rm MBE}$ roughly tracks $t_{\rm ff}(n(r))/\sqrt{r/a(r)}$, where $a(r)$ is the gravitational acceleration less the pressure gradient, \refeq{vprof}, so that $\kappa_{\rm MBE}$ can be intuitively understood as the ratio of the uniform-density free-fall timescale to the local infall timescale.}

\begin{figure}
    \centering
    \includegraphics[width=\columnwidth]{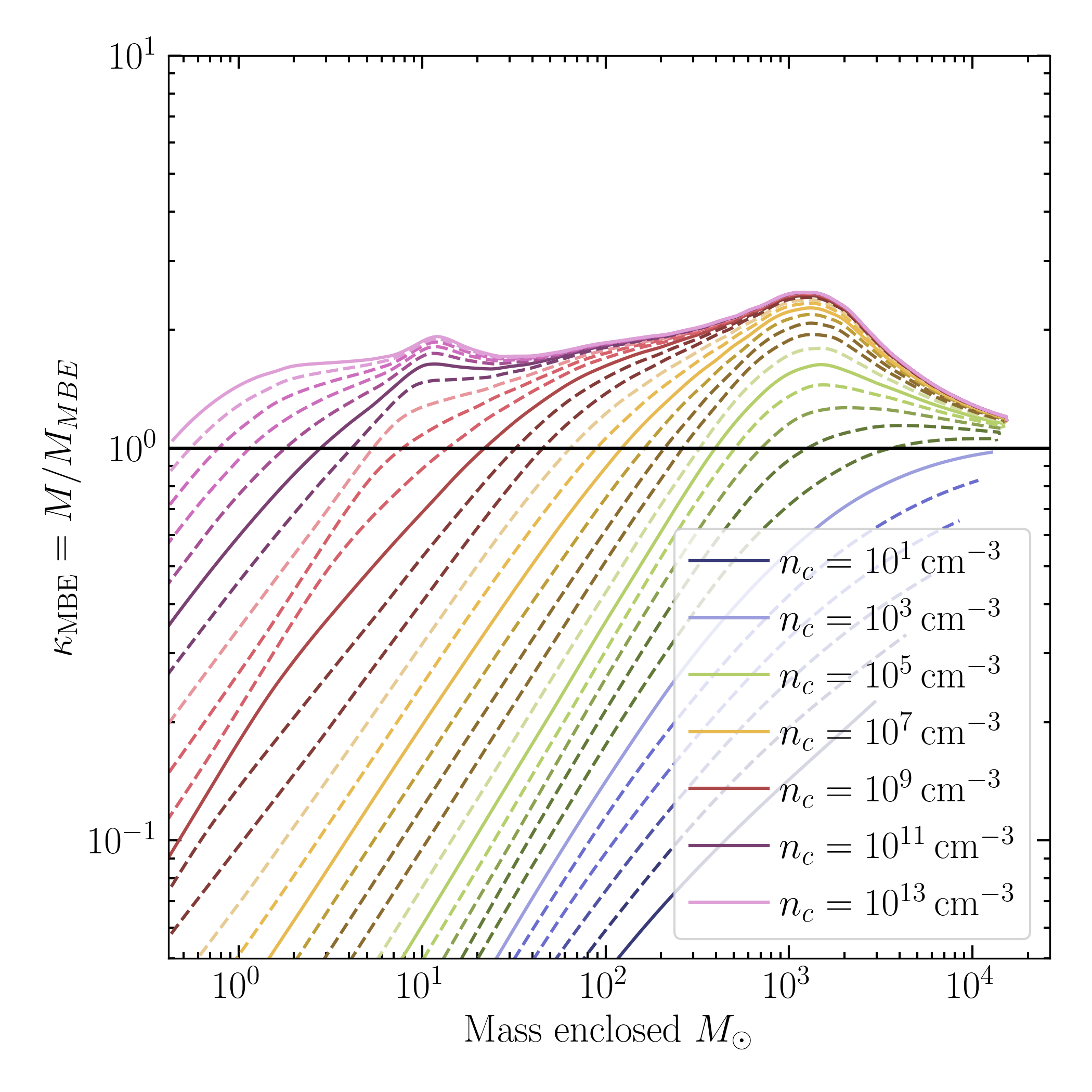}
    \vspace{-20pt}
    \caption{The ratio $\kappa_{\rm MBE}$ of mass enclosed to Bonnor-Ebert mass for a sequence of central densities, in the molecular cooling mini-halo. Gravitational instability begins to occur with the central density near the critical density of molecular hydrogen cooling ($n_c=1000 \, \rm cm^{-3}$), and $\kappa_{\rm MBE}$ peaks with $1000\, \rm M_\odot$ enclosed, corresponding to the Jeans mass at the loitering phase.}
    \label{fig:h2be}
\end{figure}

The ratio $\kappa_{\rm MBE}$ in the molecular hydrogen cooled cloud is shown in Fig.~\ref{fig:h2be} for a range of central densities. We find that gravitational instability in the envelope sets in when the central density is near the molecular cooling critical density, $n_{\rm H} \sim 10^3 \, \rm cm^{-3}$, and peaks around $1000 \, \rm M_\odot$ (i.e.~around the loitering point/mass). We also point out that Fig.~\ref{fig:h2be} differs qualitatively from similar plots in the literature based on the isothermal Jeans mass (see Appendix \ref{app:mjratio}). 

These results clarify the lore that a decreasing temperature (with density) ``promotes fragmentation'' while an increasing temperature ``suppresses fragmentation'' (e.g.~\citealt{Li_2003}). While we do not study the multiplicity of cores, our results illustrate how the characteristic mass of collapsing clouds depends on the density-temperature relationship. As illustrated by the light blue/dark green lines in Fig. 3, we see that once $\kappa_{\rm MBE}$ exceeds unity, a positive temperature gradient (i.e.~strong cooling) leads to a shallow density profile (for $r\gtrsim1~{\rm pc}$) so that $\kappa_{\rm MBE}$ hardly increases in the envelope\footnote{The maximum of $\kappa_{\rm MBE}$ also increases in the range   $10\,{\rm cm^{-3}} \lesssim n \lesssim\, 10^3 \,{\rm cm^{-3}}$ despite the positive temperature gradient because $\kappa_{\rm MBE}$ has not yet exceeded unity. In this phase, an initial Bonnor-Ebert mass of gas is accumulating in the core.}. If the cooling were to continue indefinitely, the final result would be an infinitesimal core surrounded by a nearly hydrostatic envelope. It is plausible that this nearly hydrostatic outer region (with $M(r)\sim 10^{4}\ \rm M_\odot$) could be vulnerable to, for example, turbulent fragmentation leading to the formation of multiple contracting cores. {In this picture, though, fragmentation is not invoked to explain the characteristic mass of these cores.} On the other hand, an isothermal or heating density/temperature relationship (negative temperature gradient) leads to a prompt increase in $\kappa_{\rm MBE}$, so that nearly all of the core mass at the density where the isothermal/heating part of the evolution begins is rapidly accelerated inwards, here within $M(r)\lesssim 10^3\ \rm M_\odot$. These points are further illustrated in the examples of the following sections. 

We emphasize again that this picture is quantitatively similar to but qualitatively distinct from the ``dynamical collapse'' investigated by e.g. \citet{Larson69,Penston1969,Foster93}. Rather than setting up an isothermal gas cloud out of dynamical equilibrium, we are tracking the evolution of the cloud between quasi-equilibrium states determined by the gas chemistry and cooling rates. Without cooling, the gas would rapidly heat and the contraction stall out. In this sense, the contraction of the core is always regulated by pressure and cooling. While an isothermal equation of state can provide a reasonable approximation to this balance between cooling and heating, our approach clarifies the picture by more accurately including the underlying thermochemistry of the collapsing gas. 

In particular we point out that calculations involving self-gravitating isothermal gases do not conserve the total energy of the system: there is an implicit energy loss rate imposed by the equation of state. Unlike realistic radiative cooling rates, the isothermal energy loss rate (which is just the opposite of the compressional heating term) is a function only of $\dot \rho$, which is why isothermal runaway collapse is initiated from rest only when the configuration is already dynamically unstable. 

{Finally, on a practical note, we point out that the modified Bonnor-Ebert scale can be assessed in two ways in simulations. First, $r_{\rm MBE}$ can be determined as the point where the density stays stationary in successive snapshots. Alternatively, the thermal evolution determined from the simulation can be used as the input in this model to determine the modified Bonnor-Ebert radius as a function of central density, and in turn $\kappa_{\rm MBE}$ (see Appendix \ref{app:sims}). }

\subsection{Infall Rate}
\label{sec:infall}
{We now explain the implications of $\kappa_{\rm MBE}>1$ for the dynamics of the gas, by using the density and velocity profiles before proto-star formation to estimate the infall rate after the contraction halts and a hydrostatic core is formed. The conceptual point is that for $\kappa_{\rm MBE} >1$, the density is larger than the hydrostatic value, and the gas is thus accelerated towards the core.} A widely adopted estimate of the infall rate (e.g.~\citealt{Hosokawa09,Li_2021}) is 
\begin{equation}
    \dot M \approx M_J/t_{\rm ff}\approx c_s^3/G . 
    \label{eq:cscubed}
\end{equation}
In the Larson-Penston solution (which represents a highly dynamical isothermal collapse), $\dot M \approx 47 c_s^3/G$ \citep{Hosokawa09}, while in the initially static Shu solution, the prefactor is very nearly unity \citep{Shu77}. Neither limit is typically attained in simulations (e.g.~\citealt{Hunter77,Foster93, Omukai2010}), where (in contrast to the Larson-Penston solution) the initially small infall velocity at large radii is relevant and (in contrast to the Shu solution) the envelope is not hydrostatic at the end of the core contraction phase. Moreover, these similarity solutions do not account for the departures from isothermality, which introduce new scales in the problem. 

Towards a calculation of the infall rate, we estimate the radial velocity profile of the gas once the core has become small and dense (near the epoch of proto-star formation) using the density profile calculated above. We model the radial velocity profile from the trajectory of a test particle moving towards the centre of the cloud, assuming that significant gravitational acceleration is sourced at the radius $r$ only once the core contracts to much smaller radii.

In this test-particle model, we approximate the acceleration field as constant in time but varying in space. The acceleration experienced at each radius $r$ is thus given as the gravitational acceleration from the late-time mass enclosed less the pressure gradient:
\begin{equation}
    \frac{dv}{dt} = -\frac{G [M_{DM}(r) + M(r)]}{r^2} - \frac{1}{\rho} \frac{dP}{d\rho}\frac{d\rho}{dr},
    \label{eq:vprof}
\end{equation}
where $M_{DM}(r)$ is the dark matter mass interior to $r$ (which is assumed not to evolve over the collapse) and $M(r)$ is the late-time mass enclosed (in this example, the mass enclosed when the central density is $10^{13}\,\rm cm^{-3}$). 

That is, using the identity $dv/dt=vdv/dr=\frac12 d(v^2)/dr$, we integrate the equation as
\begin{equation}
\begin{split}
&v^2(r) - v^2(r_i) \\
&\quad = 2 \int_{r_i}^{r}  dr 
\left[-\frac{G [M_{DM}(r) + M(r)]}{r^2} - \frac{1}{\rho} \frac{dP}{d\rho}\frac{d\rho}{dr}\right]\,.
\end{split}
\end{equation}

We do not model in detail the drop-off of the infall velocity near the core. Instead, we {truncate the velocity profile at 25 times the Bonnor-Ebert radius} at the highest central density in our calculation.  We impose a zero-velocity boundary condition and begin the integration when the mass enclosed first exceeds the Bonnor-Ebert mass. The assumption that the gas is accelerated from near rest is most reasonable if there is an initial quasi-static period, for example as coolants accumulate.\footnote{{It is possible to estimate a non-zero initial velocity at the Bonnor-Ebert radius by considering virial equilibrium and/or dissipation of turbulent velocity of the cloud at larger/halo scales \citep{McKee2003,Luo2024,Luo2024mc} Here, we adopt the simple zero-velocity condition in line with the canonical picture of monolithic collapse of primordial gas under inefficient cooling and weak turbulence \citep{Chon_2021}. It is shown in Appendix~\ref{app:sims} that this assumption has minor effects on the mass scale and dynamics of collapse.}}. However, we have checked that the results in the inner region are insensitive to this assumption (see App.~\ref{app:dmprofile}). Therefore, we are justified in beginning the integration at the radius $R_0$ where the following condition is satisfied: 
\begin{equation}
    t_{\rm in} = \sqrt{-R_0/a(R_0)} < 10^{7} \, \rm yr,
\end{equation}
with the acceleration $a(R_0)$ given by the right hand side of \refeq{vprof}. By this condition we avoid the situation that immediately after exceeding the Bonnor-Ebert mass \refeq{vprof} can be very stiff. 

\begin{figure}
    \centering
    \includegraphics[width=\columnwidth]{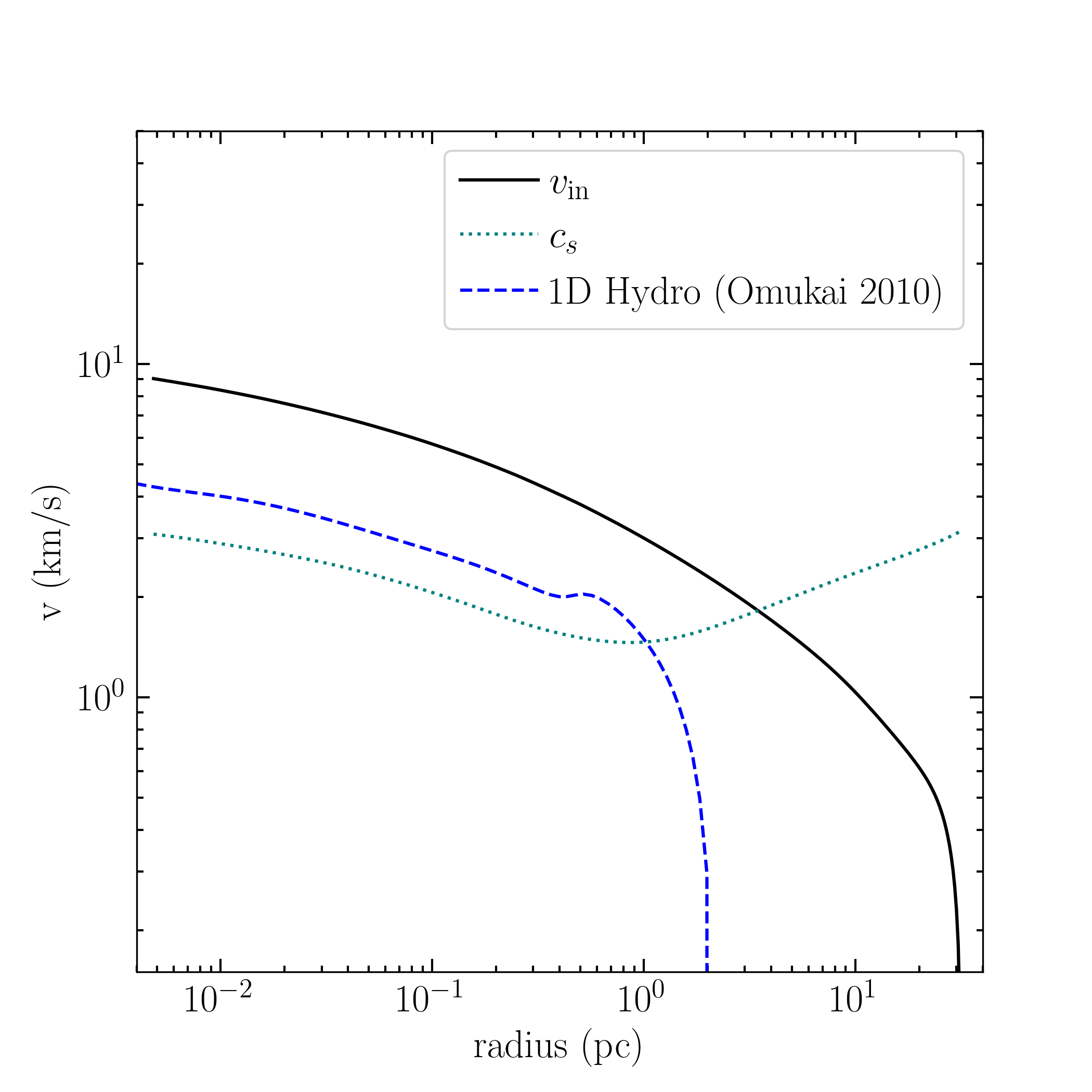}
    \vspace{-20pt}
    \caption{The radial velocity profile for the molecular cooling mini-halo (black), along with the sound speed (teal, dotted) and the velocity profile found by \protect \citet{Omukai2010} (blue, dashed). The disagreement at large radii is due to the imposition of a zero-velocity boundary condition around 2 parsecs in \protect\citet{Omukai2010}. However, a factor of (almost exactly) two discrepancy persists in the inner region.}
    \label{fig:h2vel}
\end{figure}

The velocity profile is shown in Fig.~\ref{fig:h2vel}. Our velocity profile initially (i.e.~at large radii) greatly exceeds the results of \citet{Omukai2010} because those authors imposed a zero velocity boundary condition at a smaller radius. In the inner region our velocity exceeds the 1D hydro results by a factor of almost exactly two, a discrepancy which persists even if we match the zero-velocity boundary condition to the hydro calculation. The disagreement can be explained by the fact that in our approximation that the gas at radius $r$ is accelerated by the ``very'' late time mass enclosed, rather than ``somewhat'' after the Bonnor-Ebert radius becomes smaller than $r$. That is, in reality the right hand side of \refeq{vprof} should evolve with time as the gas at $r$ is accelerated over a window of times/central densities after the core has receded from $r$ but before the central timescale becomes too short to meaningfully affect the scale $r$.

Proceeding, we construct the infall rate as 
\begin{equation}
    \dot M(r) = 4\pi r^2 \rho v,
\end{equation}
with $v$ as calculated above and $\rho$ the late-time density profile. 

\begin{figure}
    \centering
    \includegraphics[width=\columnwidth]{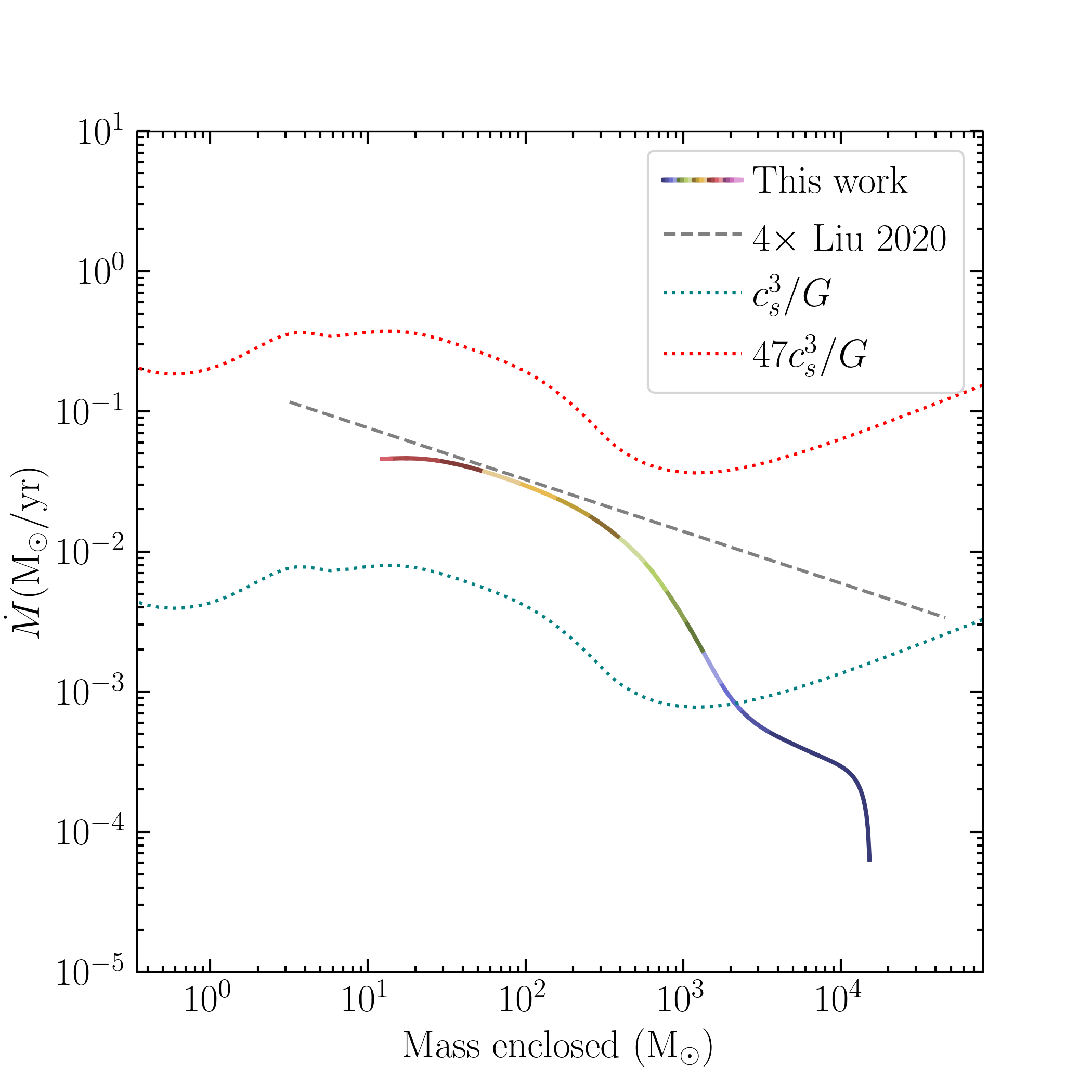}
    \vspace{-20pt}
    \caption{The infall rate as a function of mass enclosed for the molecular cooling mini-halo (solid, multi-colored), along with the Shu quasi-static accretion rate (teal, dotted), the Larson-Penston accretion rate (red, dotted), and quadruple the semi-empirical fit of the proto-stellar accretion rate from \protect \citet{Liu2020} (grey, dashed). The color is related to the mass through $M(r(n))$ in the late-time density profile, with the mapping from color to density as in Fig.~\ref{fig:phaseh2}. Note the pronounced dropoff in the accretion rate past the loitering point. For the accretion rates appropriate to the similarity solutions, the position on the horizontal axis is calculated as the Jeans mass at each $(n,T)$. }
    \label{fig:h2acc}
\end{figure}

The result is shown in Fig.~\ref{fig:h2acc}, {where we have adopted the late-time enclosed mass as the independent variable via $M = 4 \pi\int dr\, r^2 \rho$. This is an estimate based on the envelope structure of the cloud scale infall rate as a function of the mass fallen onto the disc. If the inefficiency of accretion of the protostar through the disc is neglected, this quantity can be interpreted directly as the proto-stellar accretion rate as a function of proto-stellar mass, as in (for example) \cite{Hosokawa09}.} The peak in $\kappa_{\rm MBE}$ (Fig.~\ref{fig:h2be}) corresponds to a regime of rapidly increasing infall rate. At small masses our calculations roughly tracks the estimate \refeq{cscubed}, with an overall enhancement sourced during the early, highly gravitationally unstable phase of the collapse of the envelope. This is also approximately consistent with the analytic calculation of \citet{Tan2004}, although in that work a free parameter of order unity (corresponding to the enhancement relative to the Shu solution) multiplies the accretion rate. Our estimate is a factor of few greater than the semi-empirical estimate of the proto-stellar accretion rate of \citet{Liu2020} in the regime where that fit was calibrated. The factor of few can be attributed to the inefficiency of accretion onto protostars through the accretion disc as compared with the cloud level infall rate, together with the factor of two overestimate of the velocity in our calculation. 

We find important qualitative differences relative to the Jeans estimate \refeq{cscubed}, related to the arguments discussed in the preceding section. Crucially, we demonstrate that the infall rate depends not only on the sound speed, but also on its gradient. A sound speed which decreases with increasing density is associated with a \textit{more} stable configuration (Fig.~\ref{fig:h2be} and accompanying text) and a correspondingly smaller infall rate. In contrast, the Jeans estimate $\dot M \sim c_s^3/G$ depends on the temperature alone.

{Now, we have obtained the relation between the infall rate $\dot{M}$, collapsed/enclosed mass $M$, and the corresponding cloud spatial ($r$) and density ($n\propto M/r^{3}$) scales, from which we can further derive their time evolution with $t(M)=\int_{0}^{M}[1/\dot{M}(M')]dM'$. This is a crucial step towards predicting the final outcome of the collapse. Hydrodynamic simulations and analytical models show that due to the angular momentum of the infalling gas and outflows, only a fraction $\eta\sim 0.25-0.75$ \citep{Matzner2000,Sakurai2016,Tanaka2017,Staff2019,Toyouchi_2022} of the collapsed mass $M$ is accreted by the protostars through a protostellar disc, whose size $R$ is correlated with the collapsed mass $M$ \citep{Tan2004,Liu2020}. As shown in a companion paper \citet{liu2024}, the scaling relations $\dot{M}(M)$, $t(M)$, and $R(M)$, can be used to calculate the final mass of Pop~III stars formed in the cloud with an analytical model that considers the balance between gas infall and disc photo-evaporation by the ionizing photons from protostars, and the limit of stellar mass placed by lifetime and instability. Assuming $\eta=0.5$ and only one protostar forms in the cloud for simplicity, applying our results to the model in \cite{liu2024} produces a final stellar mass of $\hat{M}_{\star}\simeq 260\ \rm M_\odot$, consistent with the results from hydrodynamic simulations of $\rm H_{2}$-cooling clouds with similar gas infall rates $\dot{M}\simeq 0.016\ \rm M_\odot\ yr^{-1}$ at the density scale of $n\simeq 10^{6}\ \rm cm^{-3}$ \citep{hirano_one_2014,Hirano2015,Toyouchi_2022,Sugimura2023}\footnote{{The original model in \citet{liu2024} assumes a power-law scaling $\dot{M}\propto M^{-0.37}$ following \citet{Liu2020}, which is valid for polynomial gas with $P\propto n^{1.09}$ \citep{Omukai1998,Tan2004}. In our case, the decline of $\dot{M}$ with $M$ is more rapid for $M\gtrsim 100\ \rm M_\odot$, likely due to the zero-velocity boundary condition (at the moment when the mass enclosed first exceeds the Bonnor-Ebert mass, see Appendix \ref{app:dmprofile}) and the deviation of the effective equation of state from a simple power-law. As a result, the final stellar mass predicted from our $\dot{M}(M)$ is lower than that from the power-law model in \citet{liu2024} by a factor of $\sim 2$. This difference is within the scatter seen in the predictions of hydrodynamic simulations.}}. }

\section{Examples}
\label{sec:ex}
We now present two additional applications of the methods developed in the preceding sections, which further elucidate the relevant physics. First, we demonstrate a case where the collapse is delayed, allowing the efficient formation of $\hd$. The resulting cooling and heating are then stronger due to the presence of $\hd$, emphasizing the arguments we have developed. Second, we present a nearly-isothermal atomic cooling example, which is in a sense the opposite extreme. 

\subsection{Delayed Core Contraction with the $\hd$ molecule}
\label{sec:delay}
\begin{figure}
    \centering
    \includegraphics[width=\columnwidth]{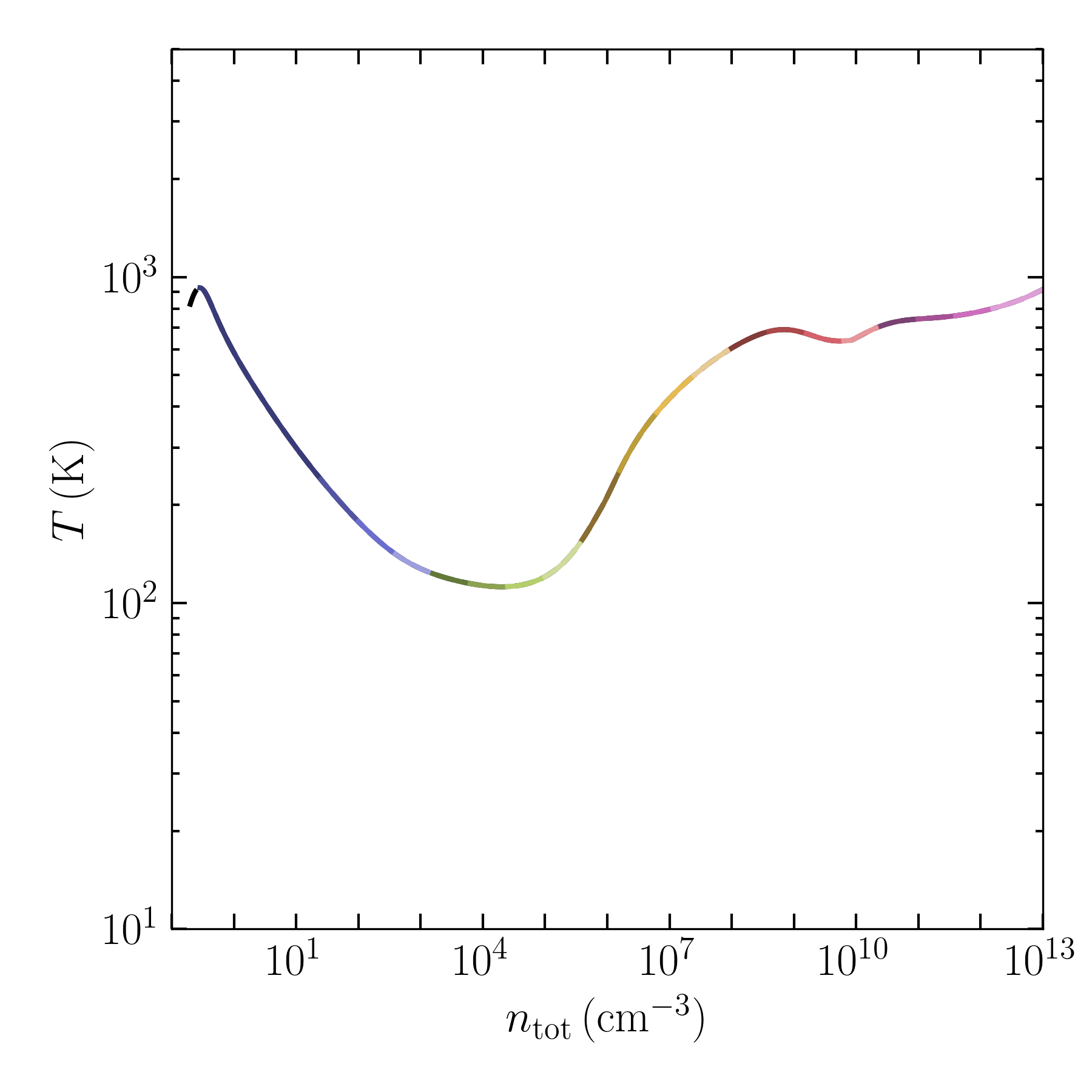}
    \vspace{-20pt}
    \caption{The temperature-density relationship for the delayed contraction, $t_{\co} = 3 t_{\rm ff}$. A lower minimum temperature and steeper temperature gradients are realized due to the production of $\hd$. }
    \label{fig:hdpd}
\end{figure}
\begin{figure}
    \centering
    \includegraphics[width=\columnwidth]{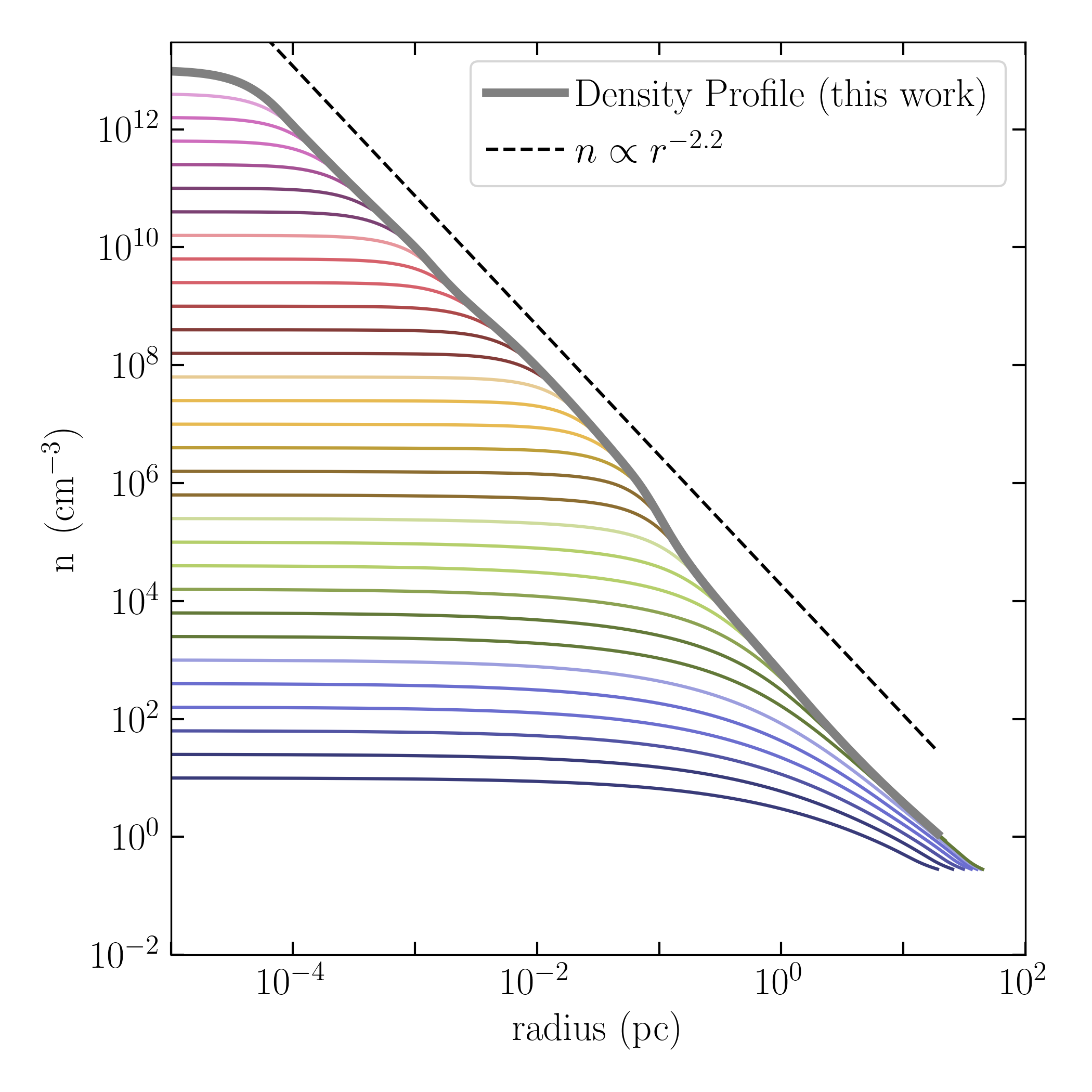}
    \vspace{-20pt}
    \caption{As Fig.~\ref{fig:h2fullprofile}, but for the delayed contraction with $\hd$ cooling. Here, lacking a reference hydro run we have simply overplotted the characteristic $n\propto r^{-2.2}$ slope. Note the more pronounced features in the density profile due to stronger deviations from isothermality. }
    \label{fig:hddens}
\end{figure}
\begin{figure}
    \centering
    \includegraphics[width=\columnwidth]{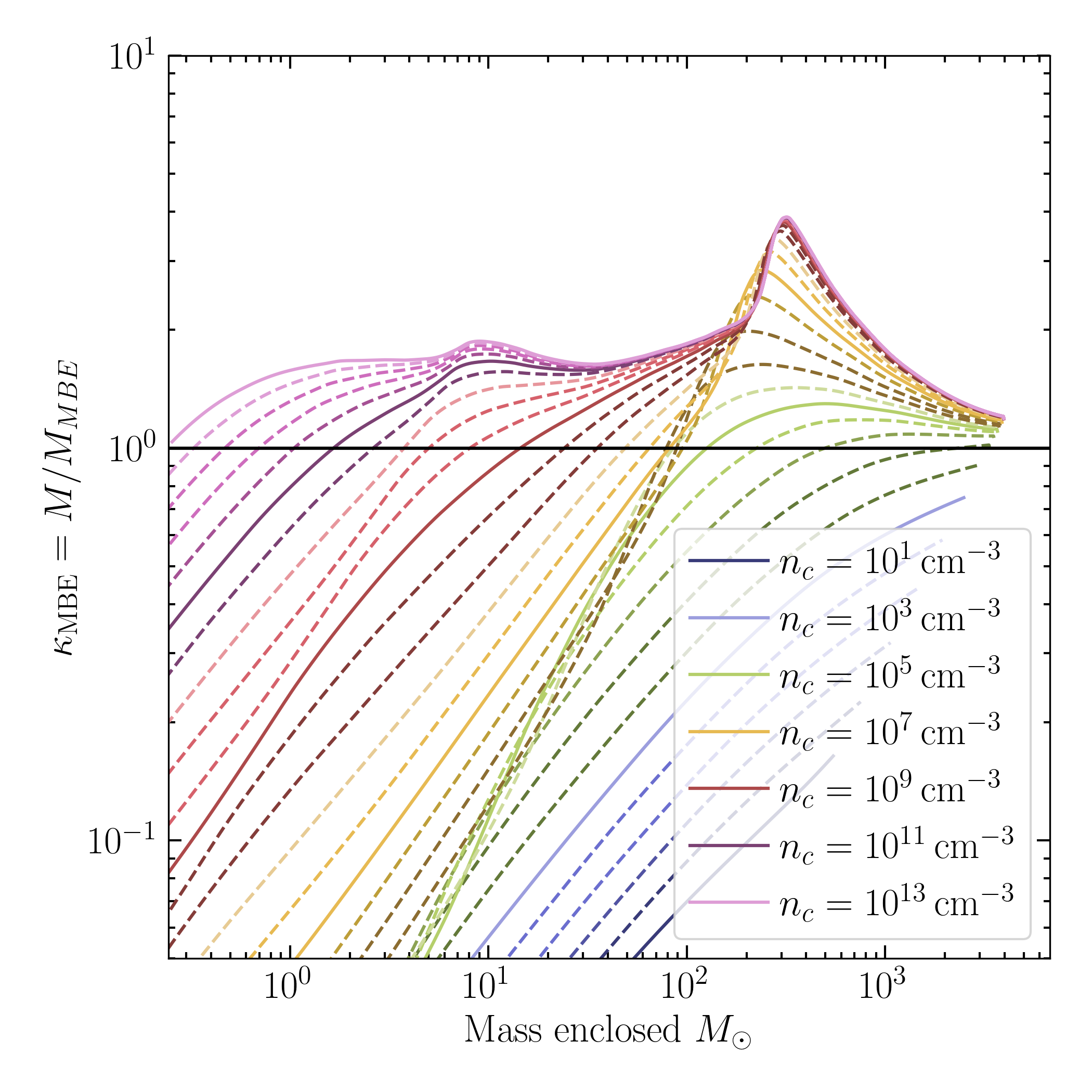}
    \vspace{-20pt}
    \caption{As Fig.~\ref{fig:h2be}, but for the delayed contraction with $\hd$ cooling. Gravitational instability in the envelope is established when the temperature begins to rise. }
    \label{fig:hdbe}
\end{figure}
If the contraction of a pristine gas cloud's core is delayed, for example, due to rotation \citep{hirano_one_2014} or an initial shortage of coolants \citep{gurian2024zero}, the chemical thermal-evolution is modified due to chemical fractionation of the $\hd$ molecule. We now apply our model to explain how this modification of the chemistry propagates into the dynamics of the collapse. To this end, we adopt initial conditions exactly as in the previous section except that we take $t_{\co} = 3t_{\rm ff}$.  It is not completely straightforward to set up a hydrodynamical simulation with realistic initial conditions which guarantee that $t_{\co} = 3t_{\rm ff}$. However, we have shown in \citet{gurian2024zero} that in the simplest case of Pop.~III star formation (neglecting, for example Lyman-Werner backgrounds, turbulence, and the baryon-dark matter streaming velocity) the delay factor can be predicted based on the host halo mass and redshift. We demonstrate in this section how such knowledge of the thermal evolution of the core can be directly extended into predictions concerning the dynamics of the collapse. Extending the calculation of the delay factor to include additional environmental factors is a target for future work.

The temperature-density relationship for this case is shown in Fig.~\ref{fig:hdpd}. Compared to the $\hto$ cooling example shown in Fig.~\ref{fig:phaseh2}, the minimum temperature here is lower, $\sim 70 \, \rm K$. Using this density-temperature relationship and \refeq{envprof}, we compute the density profile shown in Fig.~\ref{fig:hddens}. Note that owing to the the steeper temperature gradients, the density profile exhibits stronger features than that of the $\hto$ cooling example shown in Fig.~\ref{fig:h2fullprofile}.

We show the ratio $\kappa_{\rm MBE}$ in Fig.~\ref{fig:hdbe}. Compared to the $\hto$ cooling halo case (Fig.~\ref{fig:h2be}), $\kappa_{\rm MBE}$ first exceeds unity only somewhat later, when the central density is around $10^{4} \, \rm cm^{-3}$. However, $\kappa_{\rm MBE}$ stays close to one until the central density increases past $10^{6} \, \rm cm^{-3}$, which is because the temperature changes only modestly between $\sim 10^3\, \rm cm^{-3}$ and $\sim 10^5\rm \, cm^{-3}$. Beyond this point, the rapidly increasing temperature causes $\kappa_{\rm MBE}$ to rapidly increase. In fact, the temperature increases sharply enough that $M_{\rm MBE}$ briefly increases with density, so that the $\kappa_{\rm MBE}$ curves cross each other in the inner region. The eventual result is a sharp peak in $\kappa_{\rm MBE}$ near $200 \, \rm M_\odot$.  

\begin{figure}
    \centering
    \includegraphics[width=\columnwidth]{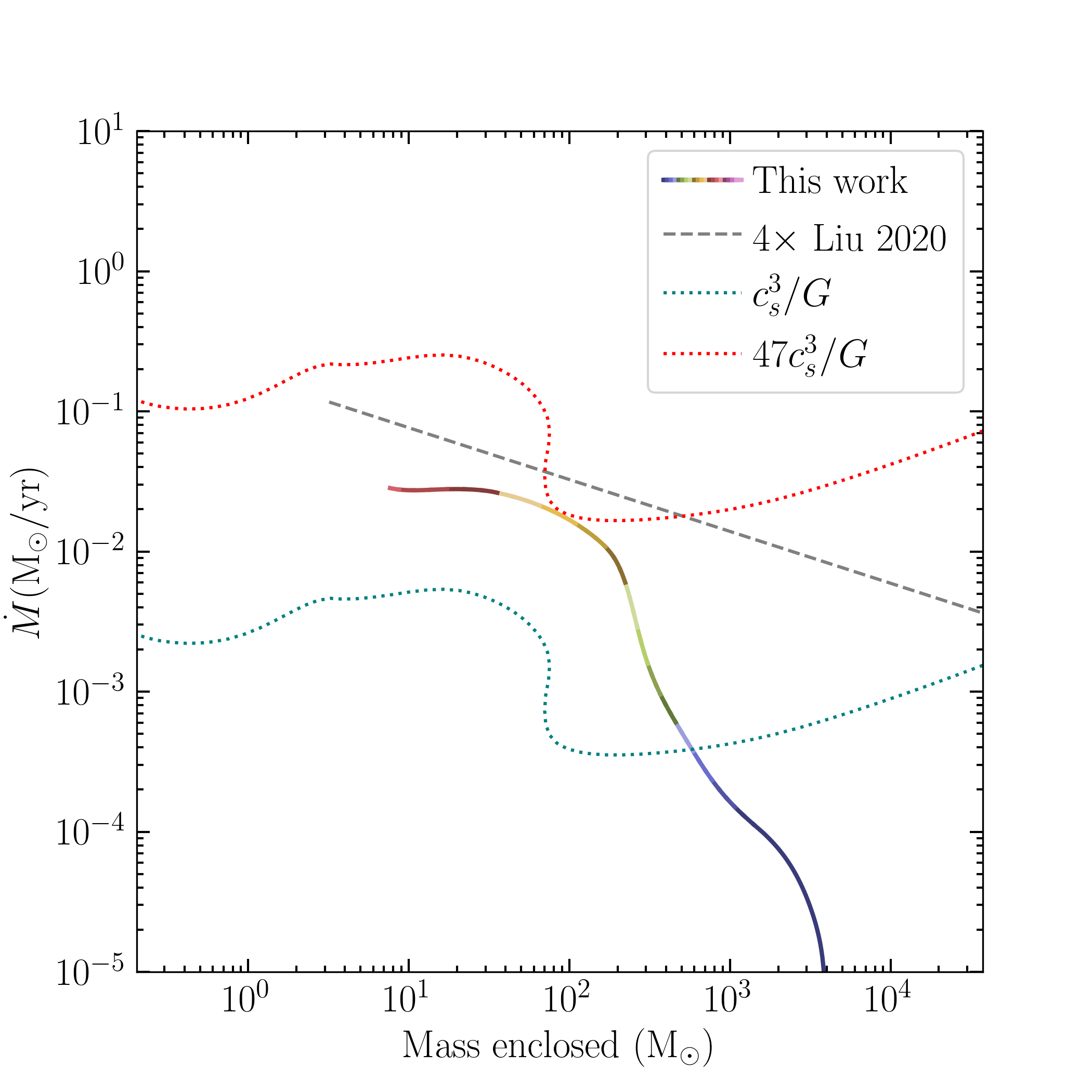}
    \vspace{-20pt}
    \caption{The accretion rate as Fig.~\ref{fig:h2acc}, but for the delayed collapse with $\hd$ cooling. Due to the strong heating once $\hd$ is destroyed, the mapping from $(n, T(n))$ to the Jeans mass (which is used to define the mass coordinate in the $c_s^3/G$ estimates) is no longer one-to-one.}
    \label{fig:hdacc}
\end{figure}

The fact that $\kappa_{\rm MBE}$ remains very close to unity for the initial part of the collapse (due to the strong cooling) means that as the central density increases the envelope remains nearly hydrostatic.  The resulting accretion rate is shown in Fig.~\ref{fig:hdacc}. In this case the $c_s^{3}/G$ estimate becomes ill-defined due to the non-monotonicity of both Bonnor-Ebert mass and Jeans mass, mentioned above. 

The results of this section are qualitatively consistent with the simulations of e.g.~\citet{hirano_one_2014, nishijima2023lowmasspopiiistar}, as discussed in detail in Appendix \ref{app:sims}. \citet{Omukai2010} also provides a benchmark for the effects of varying thermal evolution on the infall rate. The same trends of positive temperature gradients (cooling) leading to decreased infall rates while heating leads to sharply increasing infall rates are seen also in that work. However, a sharp dropoff in the infall rate is seen only at the zero-velocity boundary condition, because in that case the initial conditions were already gravitationally unstable. 
{As expected, the infall rate is lower ($\dot{M}\simeq 0.0092\ \rm M_\odot\ yr^{-1}$ at $n\simeq 10^{6}\ \rm cm^{-3}$) compared with the standard $\rm H_2$-cooling case shown in Fig.~\ref{fig:h2acc}, resulting in a lower final stellar mass $\hat{M}_{\star}\simeq 120\ \rm M_\odot$ according to the analytical model in \citet{liu2024} .}

\subsection{Atomic Cooling Halo}
\label{sec:atomic}
\begin{figure}
    \centering
    \includegraphics[width=\columnwidth]{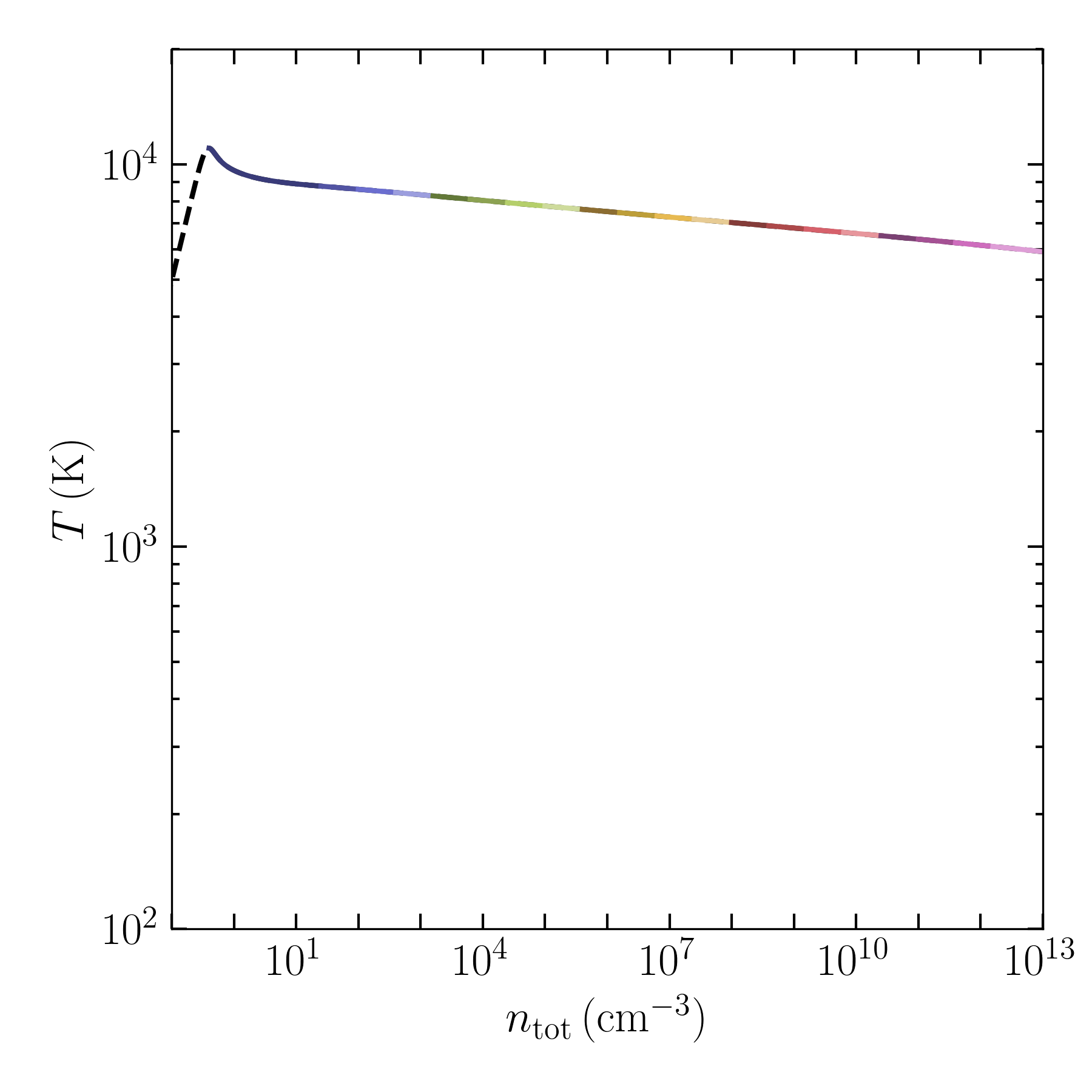}
    \vspace{-20pt}
    \caption{As Fig.~\ref{fig:phaseh2}, but for the atomic cooling halo. Here, the trajectory is nearly isothermal at the atomic cooling limit temperature.}
    \label{fig:uvpd}
\end{figure}
\begin{figure}
    \centering
    \includegraphics[width=\columnwidth]{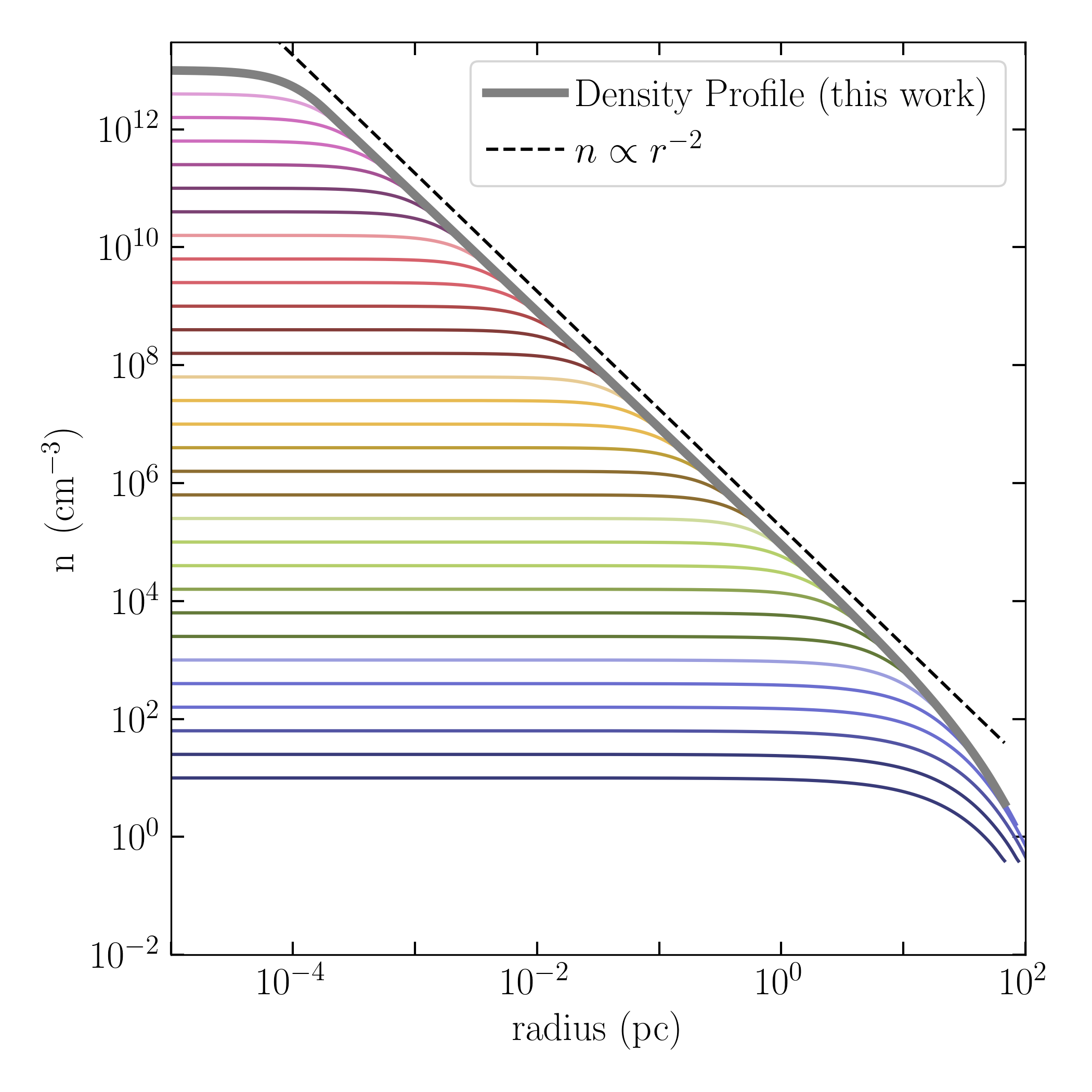}
    \vspace{-20pt}
    \caption{As Fig.~\ref{fig:h2fullprofile} but for the atomic cooling halo. From the nearly-isothermal evolution we expect $n\propto r^{-2}$, which is observed until the dark matter dominates the density at large radii.}
    \label{fig:nprofuv}
\end{figure}
\begin{figure}
    \centering
    \includegraphics[width=\columnwidth]{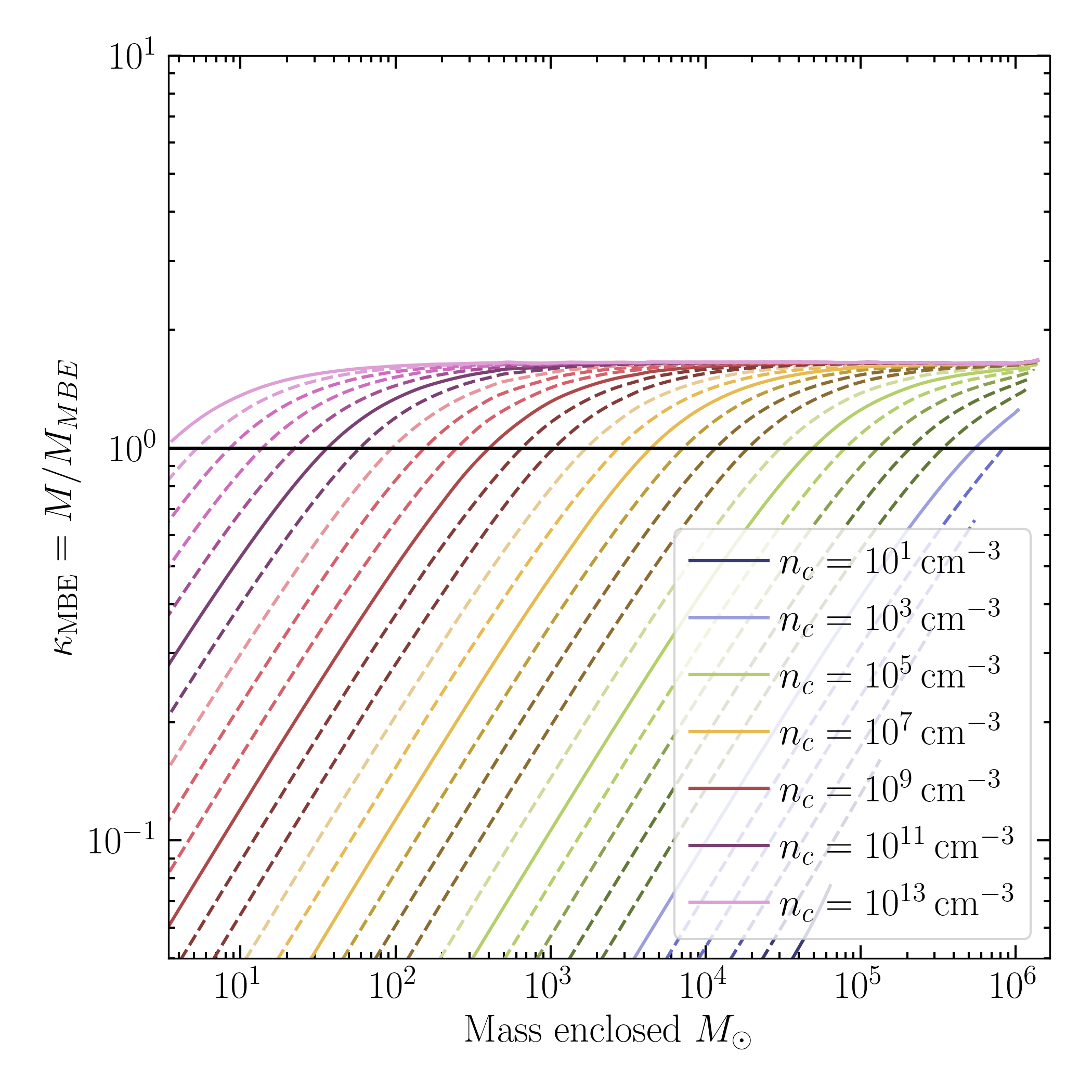}
    \vspace{-20pt}
    \caption{As Fig.~\ref{fig:h2be}, but for the atomic cooling halo. Here, an equal degree of gravitational instability is established out to the initial core mass (i.e.~the mass at which $\kappa_{\rm MBE}$ first exceeds unity).}
    \label{fig:lwbe}
\end{figure}
\begin{figure}
    \centering
    \includegraphics[width=\columnwidth]{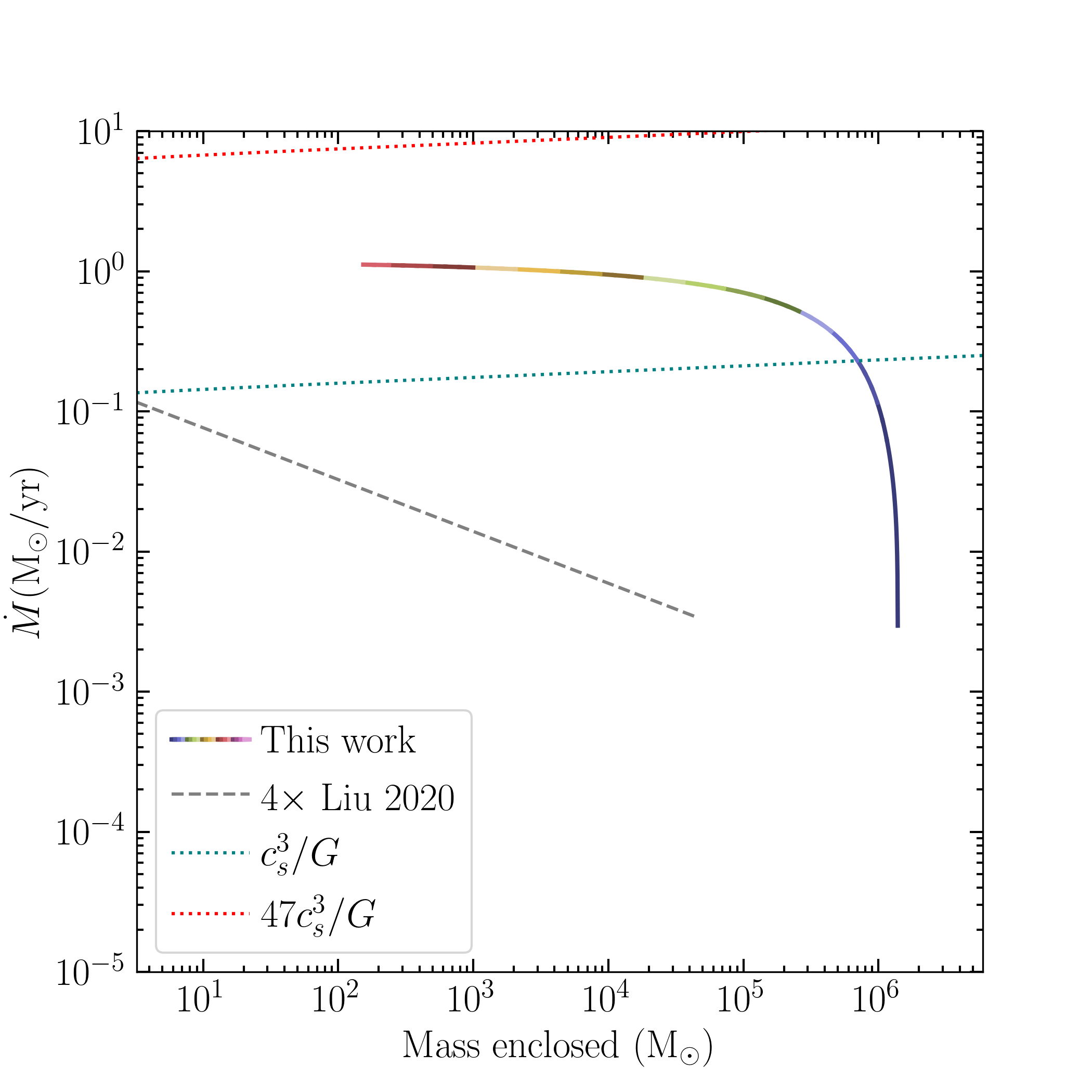}
    \vspace{-20pt}
    \caption{As Fig.~\ref{fig:h2acc}, but for the atomic cooling halo. In contrast to the ``shoulder'' seen in previous sections (associated with the temperature minimum of the gas) we see an abrupt cutoff in the infall rate beyond the initial core mass. }
    \label{fig:lwinf}
\end{figure}

If the formation of molecular hydrogen is inhibited (for example by dynamical heating due to frequent mergers, collisional dissociation, or a strong Lyman-Werner background \citep{Omukai01, Latif_2013,Wise_2019,Kiyuna_2023} a mini-halo can grow and heat up until atomic line cooling becomes efficient. This scenario can lead to the formation of supermassive ($\gtrsim 10^4 \, \rm M_\odot$) primordial stars, which may become the seeds of supermassive black holes \citep{BrommLoeb2003,Chon2018,Chon_2020, Sakurai2020,Toyouchi_2022,Reinoso_2023,Regan_2023}. However, the intrinsically large dynamic range of the problem (which depends on initial conditions for the collapse which are cosmologically rare) complicates forecasting the abundance of such objects. Here, we generate a typical atomic-cooling density-temperature relationship (Fig.~\ref{fig:uvpd}) using the \texttt{collapseUV} test provided with \textsc{krome}, where the cloud is subject to a Lyman-Werner background $J_{21} = 10^5$, with $J_{21}= J_{\rm LW}/(10^{-21} \, \rm erg\, s^{-1}\, cm^{-2}\, Hz^{-1})$. 

In this case, we assume an NFW profile for a $10^{8} \, \rm M_{\odot}$ halo with a concentration parameter $c=3.3$ (based on the mass concentration relationship of \citealt{Diemer_2015}), with the normalization calculated by the \textsc{colossus} package \citep{Diemer18}. The gas density profile is shown in Fig.~\ref{fig:nprofuv}. We find that (consistent with the nearly-isothermal evolution) the gas density scales as the inverse square of the radius, until the dark matter becomes important in the calculation of the Bonnor-Ebert radius, around $20 \, \rm pc$. A less concentrated or lower mass dark matter halo would diminish this effect. 

We show $\kappa_{\rm MBE}$ in Fig.~\ref{fig:lwbe}. The nearly isothermal evolution rapidly establishes a dynamical collapse out to the mass scale where the core contraction began (Fig.~\ref{fig:lwinf}). The characteristic (nearly constant) infall rate $\sim 1 \, \rm M_\odot/yr $ is comparable to that found in the 3D simulations of \citet{Latif_2013}, which is a case with similar thermal evolution. In the absence of strong features in the density-temperature relationship, the cloud mass is set by the mass where cooling first becomes efficient. This depends on the growth history of the halo, both through the dark matter profile (which helps set the mass enclosed at fixed density early in the collapse) and through the dynamical heating of the gas (which will determine the density at which the gas first reaches the atomic cooling limit temperature $\sim 10^4 \, \rm K$). {Given such high infall rates, stellar feedback is expected to be suppressed as the protostar will expand significantly to enter a bloating phase under rapid accretion \citep{OmukaiPalla2001,OmukaiPalla2003,Hosokawa2013,Haemmerle2018,Herrington2023,Nandal2023}. Therefore, combining our infall history with the model in \citet{liu2024}, we predict that a supermassive star of $\hat{M}_\star\sim 6\times 10^{4}\ \rm M_\odot$ will form in the end, which is expected to collapse directly into a massive black hole seed.}

\section{Discussion}

We have developed a model of gravitational collapse regulated by radiative cooling. We have illustrated how the microphysics of the gas control the density and velocity profiles established over the course of the collapse as well as the infall rate. Further, we have presented a newly general and physically precise notion of gravitational instability in this context based on the modified Bonnor-Ebert scale. We have demonstrated the agreement of our results with vastly more sophisticated numerical treatments. Our approach is computationally expedient:  generating the full, late time density profile using a grid of 60 central densities takes on the order of a few seconds on consumer hardware, which is dominated by the compile time. With the code pre-compiled and the density-temperature relationship pre-computed, generating the density profile requires only $\sim 0.1$ seconds. In certain situations this speedup compared to hydrodynamical simulations (in exchange for some loss of accuracy) may be useful. 

However, our model does not capture the full degree of complexity present in hydrodynamical simulations (let alone reality). For example, \citet{Omukai2010} found in their one-dimensional simulations that strong heating in the core leads to the formation of shocks as the core fails to ``stay ahead'' of the infalling material. We have further made no attempt to model phenomena including deviations from spherical symmetry (which can lead to the formation and subsequent fragmentation of an accretion disc), turbulence, magnetic fields, and radiative feedback\textemdash {all of which are understood to play important roles in the star formation process \citep[e.g.,][]{Larson1973,McKee2002,McKee2003,Tan2004,Hennebelle2008,Hopkins2012,Guszejnov2015,Tsukamoto2015,Inoue2020,Kimura2021,liu2024,Luo2024,Thomasson2024}}. Some of these shortcomings can be addressed by future work, for example by the inclusion of additional pressure terms in an effective sound speed. 

These caveats do not diminish the utility of our model both as a cross-check for simulations in varying physical environments and as a conceptual framework. We have made precise the sense in which the density-temperature relationship in the core controls the dynamics of the entire collapse, and determines the mass of the eventual collapsing cloud. {We clearly distinguish the roles of Rees-Ostriker and Bonnor-Ebert instability criteria in the collapse of thermally supported gas. The distinction is based on the ``two-phase'', non-homologous nature of the collapse in which the gas first contracts to proto-stellar densities and then subsequently falls onto the nascent proto-star (or its disc). The Rees-Ostriker criterion controls the onset of runaway contraction in the sense of a gas core rapidly condensing to high density.} This phase of the collapse, although it can occur on a dynamical timescale, is a quasi-equilibrium process.  

This runaway cooling in the core is the cause of gravitational instability, rather than the consequence. {Then ``dynamical'' or Bonnor-Ebert instability impacts the dynamics principally after the formation of the proto-star, during the accretion phase.} {These facts are not widely appreciated. For example, it is common practice to begin cloud-scale simulations of primordial star formation with the density enhanced relative to the hydrostatic value by some constant factor to ``initiate the collapse'' (e.g.~\citealt{Omukai2010, Chon_2021}). In fact, in \cite{Omukai2010} by the time of the first snapshot the density profile has ``corrected'' to the one calculated in this work (Fig.~\ref{fig:h2fullprofile})!}  

Finally, we illustrated that the mass scale at which the core contraction initiates dynamical instability in the envelope depends crucially on the features in the density temperature relationship: strong cooling leads to stability and mild cooling or heating leads to instability. With these insights, we can make newly precise statements about the effects of the gas equation of state on the mass scale of the collapse. For example, we have extended the conventional wisdom that a nearly isothermal equation of state (as in our atomic cooling example) ``suppresses fragmentation'' \citep{Li_2003} by showing (Fig.~\ref{fig:lwbe}) that a nearly-isothermal equation of state rapidly establishes gravitational instability at the scale where the core contraction begins, which may lead to a monolithic collapse at this scale. On the other hand, compared to the argument that cooling promotes hierarchical fragmentation down to the temperature minimum, we have shown that strong cooling (as in our delayed collapse example) leads to a nearly hydrostatic envelope, so that a large infall velocity is established only past the temperature minimum. This is true independent of the possible multiplicity of the cores\textemdash fragmentation may occur in the envelope, but is not necessary to explain the characteristic mass of collapsing clouds.

By these arguments, we clarify the significance of the ``loitering point'' in Population III star formation: the increase in temperature and corresponding steep density profile at densities above the loitering point accelerates the envelope inwards, so that the characteristic mass of the collapsing cloud corresponds to the Bonnor-Ebert mass at this point. 
We show in App.~\ref{app:dmprofile} that the effect of the thermal evolution on the dynamics is exaggerated by the fact that dark matter dominates the potential at densities below the loitering point, further suppressing gravitational instability at low densities.  

These ideas differ from pre-existing notions concerning gravitational collapse and fragmentation based on perturbative instabilities in the medium. Because density perturbations grow on the free-fall timescale, such instabilities are not likely to operate during free-fall core contraction without external forces. Such instabilities become important, for example, when the collapse is delayed (i.e.~by inefficient angular momentum transport, resulting in the formation of a disc) or when large density perturbations are established on sub-dynamical timescales (i.e.~by supersonic turbulence). Either or both effect can easily be relevant in realistic situations. Here, we have illustrated the sense in which even a monolithic collapse contains a preferred mass scale dictated by the radiative physics of the gas. 

Our model describes the density profile of gas after runaway Kelvin-Helmholtz contraction is initiated. For a fixed density-temperature relationship, the late collapse results are fairly insensitive to the initial conditions. In Sections \ref{sec:delay} and \ref{sec:atomic} we have developed two representative examples where the cloud/halo scale physics significantly alter the density-temperature relationship, and hence the dynamics of the collapse. In the primordial case considered here, the large scale initial conditions are dictated by cosmology, and in particular by the distribution of dark matter. In a companion paper \citep{liu2024} we develop a model relating the cloud-scale infall rate (as calculated here) with the final stellar mass based on the interplay between radiative feedback and fragmentation, while in \citet{gurian2024zero} we predicted the chemical-thermal evolution of the cloud based on the cosmological environment. 
{These efforts can be connected towards a comprehensive analytic model of primordial star formation. 
This model is able to predict the final mass of stars formed from any input evolution track of primordial star-forming gas in the temperature-density phase diagram with minimal computational cost and physically motivated free parameters (for disc fragmentation, stellar evolution and feedback) rather than phenomenological parameters such as star formation efficiency. It covers all possible modes of Pop~III star formation known to date, as illustrated here using the evolution tracks predicted by the one-zone approach for the three examples of $\rm H_2$- (Sec.~\ref{section:GC}), HD- (Sec.~\ref{sec:delay}), and atomic-cooling (Sec.~\ref{sec:atomic}) clouds \citep[see also][]{liu2024}. The universality and flexibility of this model offer promising applications across a wide range of topics. For example, it can be employed to investigate the gravitational, thermal, and chemical impacts of DM physics on Pop III star formation. It can also be incorporated into cosmological simulations and semi-analytical models that account for the large-scale environmental effects (e.g., radiation background of $\rm H_2$-dissociating photons, dynamical heating by halo mergers, streaming motion between DM and baryons) on (the onset of) Pop~III star formation but lack the resolution to fully follow the small-scale cloud collapse and protostar formation/evolution processes. As long as the initial collapse of the cloud at a density scale of $n\sim 10^{3-5}\ \rm cm^{-3}$ is captured in such large-scale models, the final product of subsequent evolution can be easily derived from our analytical model. Exploring such applications is an intriguing direction for future research.}

\section*{Acknowledgements}
We thank Kazuyuki Omukai for sharing the data from \cite{Omukai2010}. We thank Chris Matzner, Sarah Shandera, and Daisuke Toyouchi for useful discussions. BL gratefully acknowledges the support of the Royal Society University Research Fellowship {and the funding from the Deutsche Forschungsgemeinschaft (DFG, German Research Foundation) under Germany's Excellence Strategy EXC 2181/1 - 390900948 (the Heidelberg STRUCTURES Excellence Cluster)}. DJ is supported by KIAS Individual Grant PG088301 at Korea Institute for Advanced Study.
NY acknowledges financial support from JSPS Research Grant 20H05847. Research at the Perimeter Institute is supported in part by the Government of Canada through the Department of Innovation, Science and Economic Development Canada and by the Province of Ontario through the Ministry of Colleges and Universities. The development of this work benefited from in-person collaboration supported by the scientific partnership between the Kavli IPMU and Perimeter Institute.

\section*{Data Availability}
The code and data underlying this paper will be shared on reasonable request to JG at \url{jgurian@perimeterinstitute.ca}. 

\bibliographystyle{mnras}
\bibliography{popiii}

\appendix
\section{The Collapse Timescale}
\label{app:fvar}
\begin{figure}
    \centering
    \includegraphics[width=\linewidth]{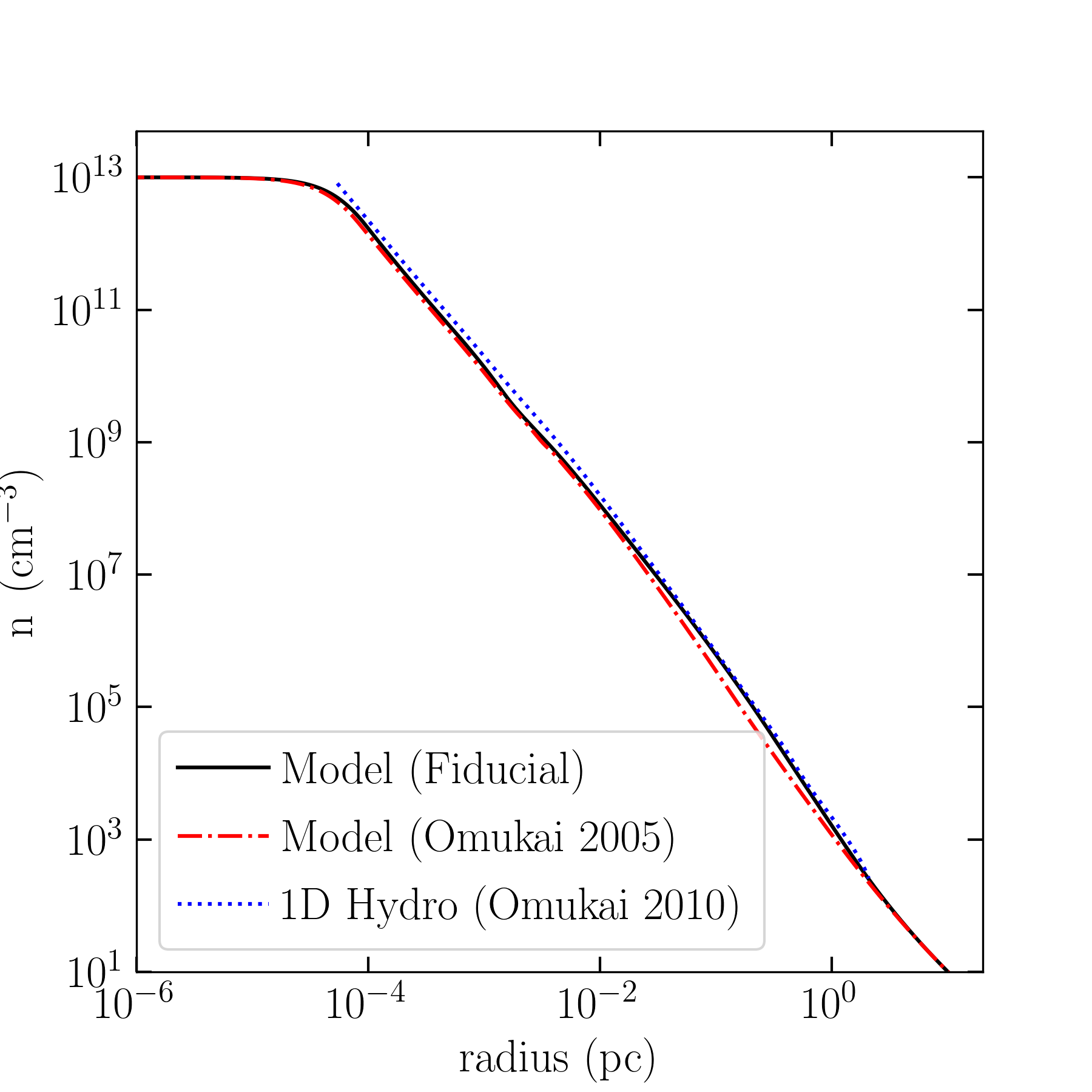}
    \caption{The density profile in the molecular-cooling cloud in our fiducial model (black), our model with the temperature-density relationship calculated as in \protect\cite{Omukai2005} (red, dot-dashed), and in the 1D hydro calculation of \protect\cite{Omukai2010} (blue, dashed). }
    \label{fig:onezcompare}
\end{figure}

{
In this work we take the collapse timescale $t_{\co}$ as a fixed, constant multiple of the freefall timescale, $t_{\co} = f t_{\rm ff}$. Other authors have considered more sophisticated one-zone models. For example, \cite{Omukai2005} adopted the following parameterization to take into account the (temporal) slowdown of the collapse due to the evolution of equation of state:
\begin{equation}
    t_{\co} = \frac{1}{\sqrt{1-f}}t_{\rm ff},
\end{equation}
where $f$ varies according the ratio of pressure to gravity, 
\begin{equation}
   f = \begin{cases}
        0 & \gamma <0.83\\
        0.6 + 2.5 (\gamma -1) - 6.0 (\gamma -1)^2 & 0.83 < \gamma <1 \\
        1.0 + 0.2 (\gamma - 4/3) - 2.9(\gamma - 4/3)^2 & \gamma >1
    \end{cases},
\end{equation}
given $\gamma = \frac{\partial \log P}{\partial \log \rho}$.
To implement this in our one-zone model, we avoid unphysical oscillations in $f$ by at each step calculating $\gamma=(\gamma_{\rm old} + \gamma_{\rm new})/2$, where $\gamma_{\rm old}$ is the value adopted at the previous timestep and $\gamma_{\rm new}$ is the value calculated at the current timestep. We show in Fig.~\ref{fig:onezcompare} that by somewhat reducing the temperature at $n\gtrsim 10^3\,\rm cm^{-3}$ due to reduced compressional heating, this alternative parameterization has a modest effect on the density profile and hence all the derived quantities in this work. The effect of this parameterization on the other cases is similarly small. However, as the examples in the main text illustrate, a global slowdown in the collapse does significantly alter the chemical-thermal evolution. In \cite{gurian2024zero} we studied cases where the global slowdown parameter can be predicted based on the cosmological environment.}

\section{The Bonnor Ebert Mass}
\label{app:be}
The usual Bonnor-Ebert criterion 
\begin{equation}
    \frac{\delta P}{\delta V} = 0 
\end{equation}
can be written in terms of the change in central density as 
\begin{equation}
    \left(\frac{\partial P}{\partial\rho_c}\right)_M\left(\frac{\partial V}{\partial \rho_c}\right)_M^{-1} = 0,
\end{equation}
where the subscript $M$ indicates the derivatives are evaluated at fixed mass. The first zero occurs when  
\begin{equation}
    \left(\frac{\partial P}{\partial \rho_c}\right)_M = 0,
\end{equation}
because the point of equal mass enclosed (where $\frac{\partial V}{\partial \rho_c}=0$) will occur at larger radius than the point of equal density. This resembles \refeq{dpdrhoc}. Using the chain rule, the Bonnor-Ebert criterion is 
\begin{equation}
    \left(\frac{\partial P}{\partial \rho_c}\right)_M = \left.\frac{\partial P}{\partial \rho}\right|_{\rho=\rho_{\rm HSE}} \left(\frac{\partial \rho_{\rm HSE}}{\partial \rho_c} +\frac{\partial \rho_{\rm HSE}}{\partial r} \frac{\partial r}{\partial \rho_c}\right)=0, 
    \label{eq:berhoc}
\end{equation}
where the second term enforces mass conservation via 
\begin{equation}
\frac{\partial r}{\partial \rho_c} =   - \frac{1}{4 \pi r^2\rho} \frac{\partial M}{\partial \rho_c}.
\label{eq:bemcons}
\end{equation}
We have checked that this formulation \refeqs{berhoc}{bemcons} agrees with Eq.~3.3 of \citet{Bonnor1956}. Clearly, the condition employed in this work \refeq{rbe} corresponds to the first term of \refeq{berhoc}, which corresponds to evaluating the derivative at fixed radius.
\section{Role of Dark Matter}
\label{app:dmprofile}
\begin{figure}
     
         \includegraphics[width=0.5\textwidth]{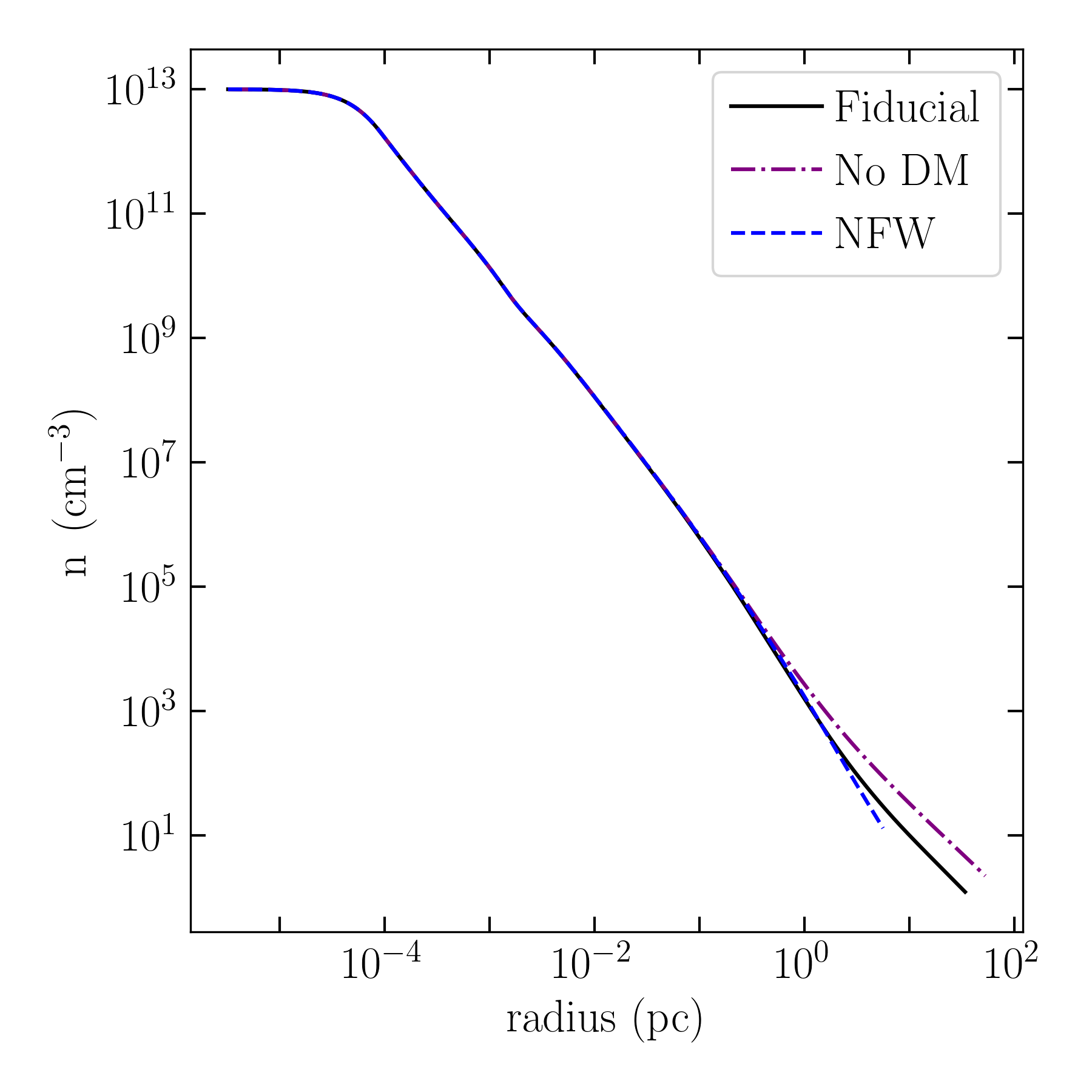}
         \vspace{-20pt}
         \caption{The dependence of the gas density profile in the molecular cooling halo on the assumed dark matter density. The black curve is when the profile is given by \refeq{dmsim}, used in the main text. The purple, dot-dashed curve represents no dark matter, while the blue, dashed curve has an NFW  DM profile, for a $5\times 10^5 \, \rm M_\odot$ halo with a concentration parameter of $c=2.8$.}
         \label{fig:dmdepprof}
    
\end{figure}

In the $\hto$ cooling mini-halo the total density is dominated by dark matter once the gas density drops below $\sim 10^3 \, \rm cm^{-3}$. Both because we do not attempt to self-consistently model the evolution of the dark matter and because the profile adopted \refeq{dmsim} is highly approximate, we here bracket the effects of our ignorance of the correct profile on our results. In addition to the fiducial profile \refeq{dmsim}, we consider both an NFW profile appropriate to a halo of mass $5\times 10^5 \, \rm M_\odot$ with a concentration parameter $c=2.8$ (i.e.~a significantly larger dark matter density than the fiducial case) and the case of no dark matter whatsoever. Adiabatic contraction \citep{Blumenthal1986} of the dark matter in response to the gas collapse can greatly enhance the dark matter density compared to any of these estimates, which may in turn have dramatic effects on the star formation process \citep{Spolyar2008}, a possibility we do not treat here. The gas density profiles in our three assumed dark matter profiles are shown in Fig.~\ref{fig:dmdepprof}. As we have already argued, the presence of (more) dark matter steepens the density profile. 

\begin{figure}
     
         \includegraphics[width=0.5\textwidth]{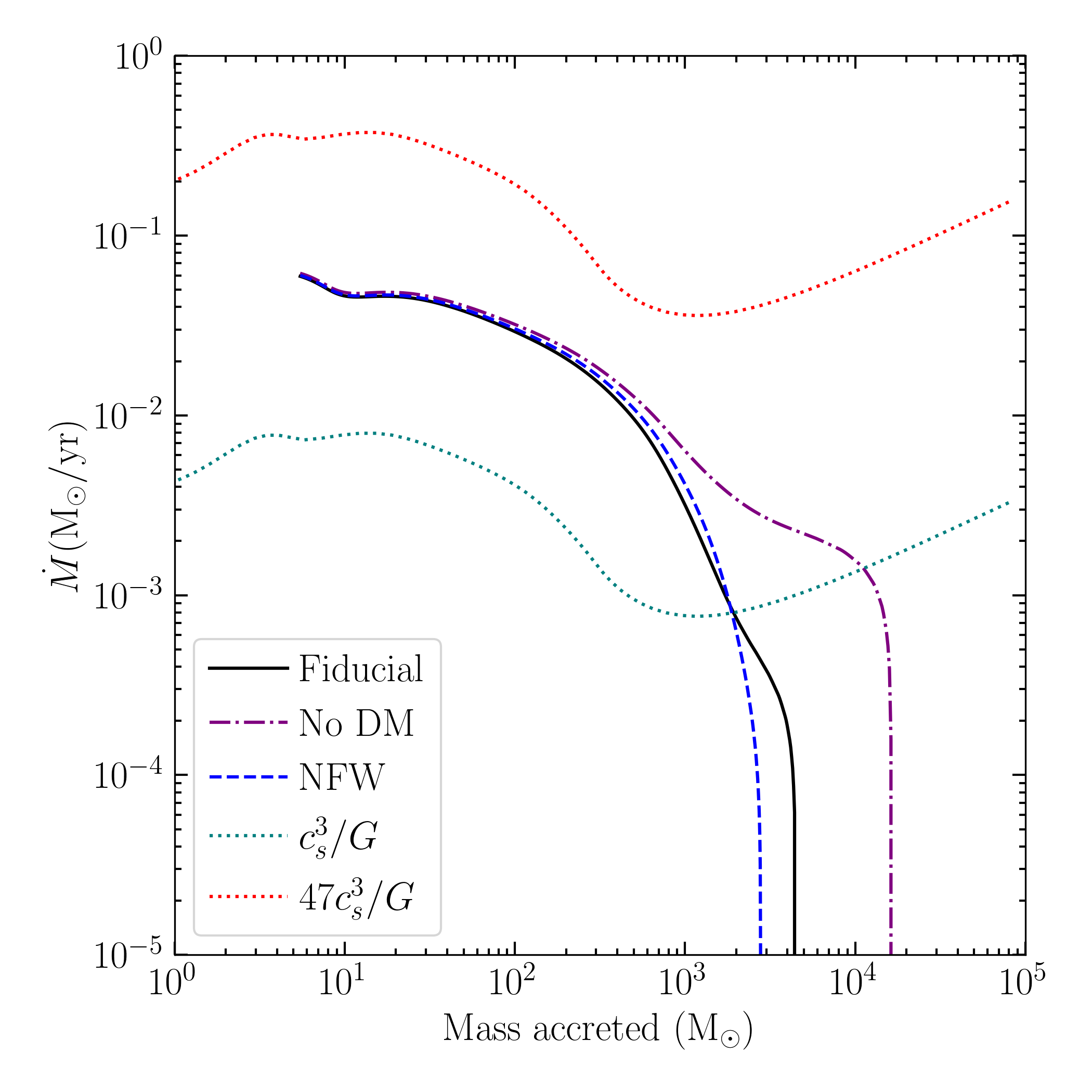}
         \vspace{-20pt}
         \caption{The accretion rate in the molecular cooling halo in the fiducial (black) case, as well as for no dark matter (purple, dot-dashed) and an NFW profile (blue, dashed). The Shu and Larson-Penston values are also shown. With no dark matter, gravitational instability extends past $10^4 \, \rm M_\odot$, albeit at a lower accretion rate.}
         \label{fig:dmdepinfall}
    
\end{figure}

The accretion rate for all three cases is shown in Fig.~\ref{fig:dmdepinfall}. In the absence of dark matter, gravitational instability sets in at a lower central density/larger mass scale, because the Bonnor-Ebert (gas) mass at a given central density is larger without dark matter contributing to the potential. A similar phenomenon is observed in simulations of pristine gas clouds separated from dark matter overdensities by supersonic streaming motions, but in that case the density-temperature relationship is additionally modified by the extreme environment \citep{nakazato_h_2022}. 

\section{The Isothermal Jeans Mass}
\label{app:mjratio}
\begin{figure}
    \centering
    \includegraphics[width=0.5\textwidth]{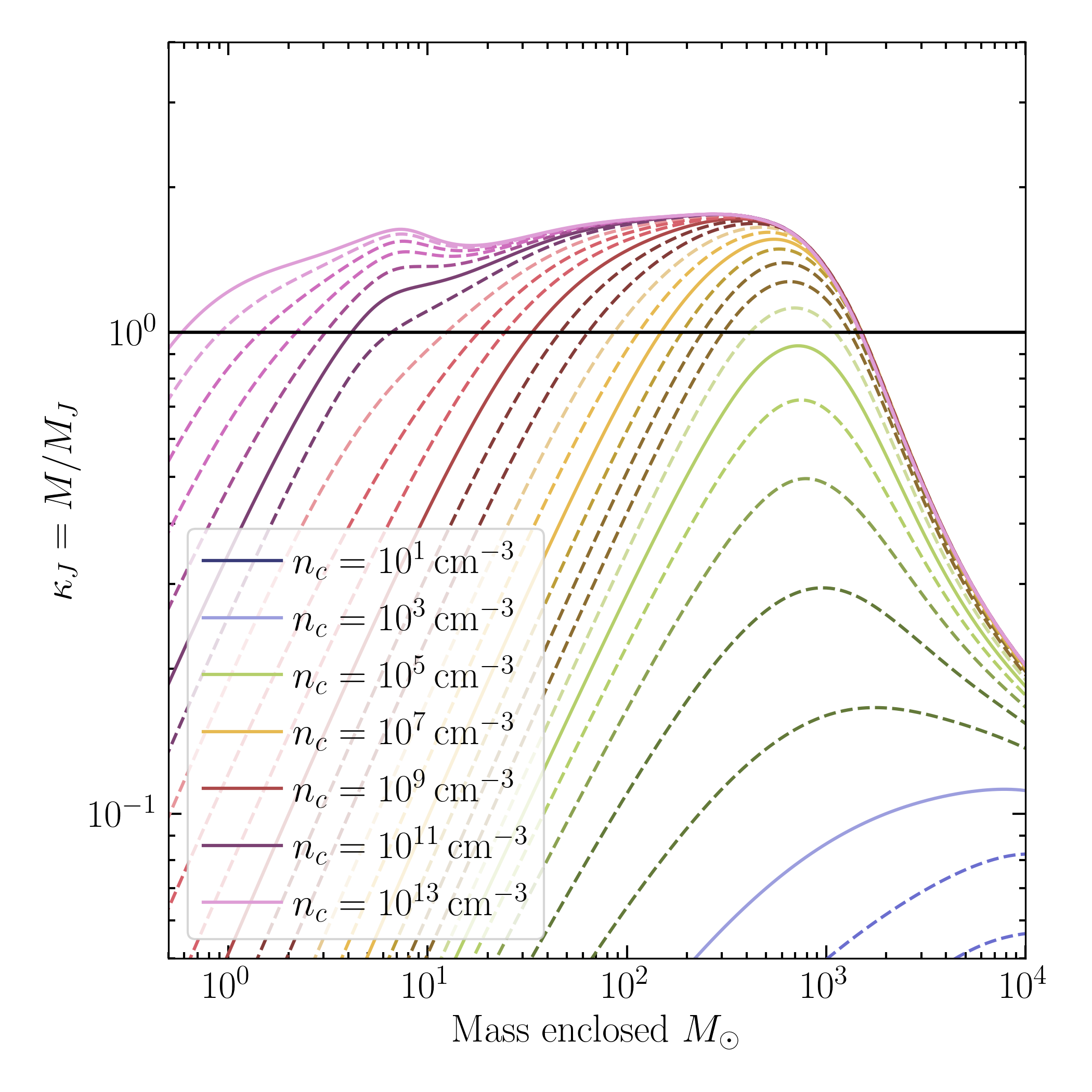}
    \vspace{-20pt}
    \caption{The ratio of the mass enclosed in the molecular cooling halo to the isothermal Jeans mass, computed from the mass-weighted temperature and mean density.}
    \label{fig:h2jeans}
\end{figure}

In the literature a quantity similar to $\kappa_{BE}$ defined here is often calculated, but instead of the modified Bonnor-Ebert mass defined here, the coefficient is calculated as the isothermal Jeans mass of the mass-weighted temperature and average density:

\begin{align}
    \bar T &= \frac{1}{M}\int_0^{M} {dM'} \,T(M')\\
    \bar \rho &= \frac{3M}{4 \pi R^{3}},
    \label{eq:rhobar}
\end{align}
so that 
\begin{equation}
    M_J \approx 1.44\left(\frac{ k_B  \bar T}{\mu m_P G }\right)^{3/2}\bar\rho^{-1/2}.
    \label{eq:mjiso}
\end{equation}
We here calculate $\kappa_{J}$ using \refeq{mjiso} in our $\hto$ cooling halo, shown in Fig.~\ref{fig:h2jeans}. The result becomes qualitatively more similar to e.g.~Fig.~13 of ~\citet{hirano_one_2014} and Fig.~2 of \citet{smith2024does} in this case. Note that for an isothermal Bonnor-Ebert sphere, \refeq{rhobar} will become small outside of the core. This should not however be interpreted as indicating a maximum mass scale for gravitational instability.  

We point out that $\bar \rho$ can also be equated with the one-zone density in \citet{smith2024does}. Such a calculation gives a qualitatively correct result without explicating the mechanism by which radiative cooling sources gravitational instability. Compared with \citet{smith2024does}, the present work does not attempt to model the initial, slow contraction during which the environmental factors establish the chemistry for the runaway collapse. Here, we have demonstrated that (absent non-thermal support) detailed modelling of the evolution of the average density is unnecessary once cooling becomes efficient. As soon as cooling kicks in, $t_{\co} \sim t_{\cc}$ and the density and mass scale at which gravitational collapse begins are already determined.

\section{Comparison With Simulations}
\label{app:sims}
{ In the main text we demonstrated reasonable agreement between the model and the spherically symmetric simulation of \cite{Omukai2010}, which adopts the same level of idealization as the model. That comparison showed that the mechanism of gravitational instability identified in this work is plausibly responsible for initiating dynamical gravitational collapse in Pop.~III star forming regions. Now, we compare our model with the 3D hydrodynamical simulations of \cite{hirano_one_2014, Sugimura2023, nishijima2023lowmasspopiiistar}. In the rest of this work, we have adopted density-temperature relationships computed using one-zone models. In fact the input for the model is a density-temperature relationship, no matter its source. In principle, one could supply the density-temperature relationship from marginally resolved scales in simulations, or by extrapolating simulation results to higher densities using one-zone models. Here, we use the density-temperature relationship from high-resolution simulation results to predict the dynamics in our model and check the extent to which the results agree with the full hydrodynamical result. For reference, we also show the appropriate comparison with the examples considered in the main body of this work. In the $\hto$ and $\hd$ cooling, we have adopted the dark matter profile \refeq{dmsim}, while for the atomic cooling cases we adopt the same NFW profile as in Section \ref{sec:atomic}. As shown in App.~\ref{app:dmprofile}, the qualitative results depend on the presence of dark matter only in the outer region. The purpose of the comparison is two-fold. On one hand, we are testing the importance of 3D effects neglected in the current model. These include small-scale (aspherical) density perturbations, turbulence, and rotation. On the other, we are assessing the accuracy of the one-zone models in determining the thermal evolution, and the effects of any inaccuracies on the dynamics of the collapse. Note, however that the one-zone models were not specifically tuned to match the simulation cases in terms of initial density, temperature, or (in the $\hd$ case) CMB temperature.  }

\begin{figure*}
      \centering
    \includegraphics[width=\textwidth]{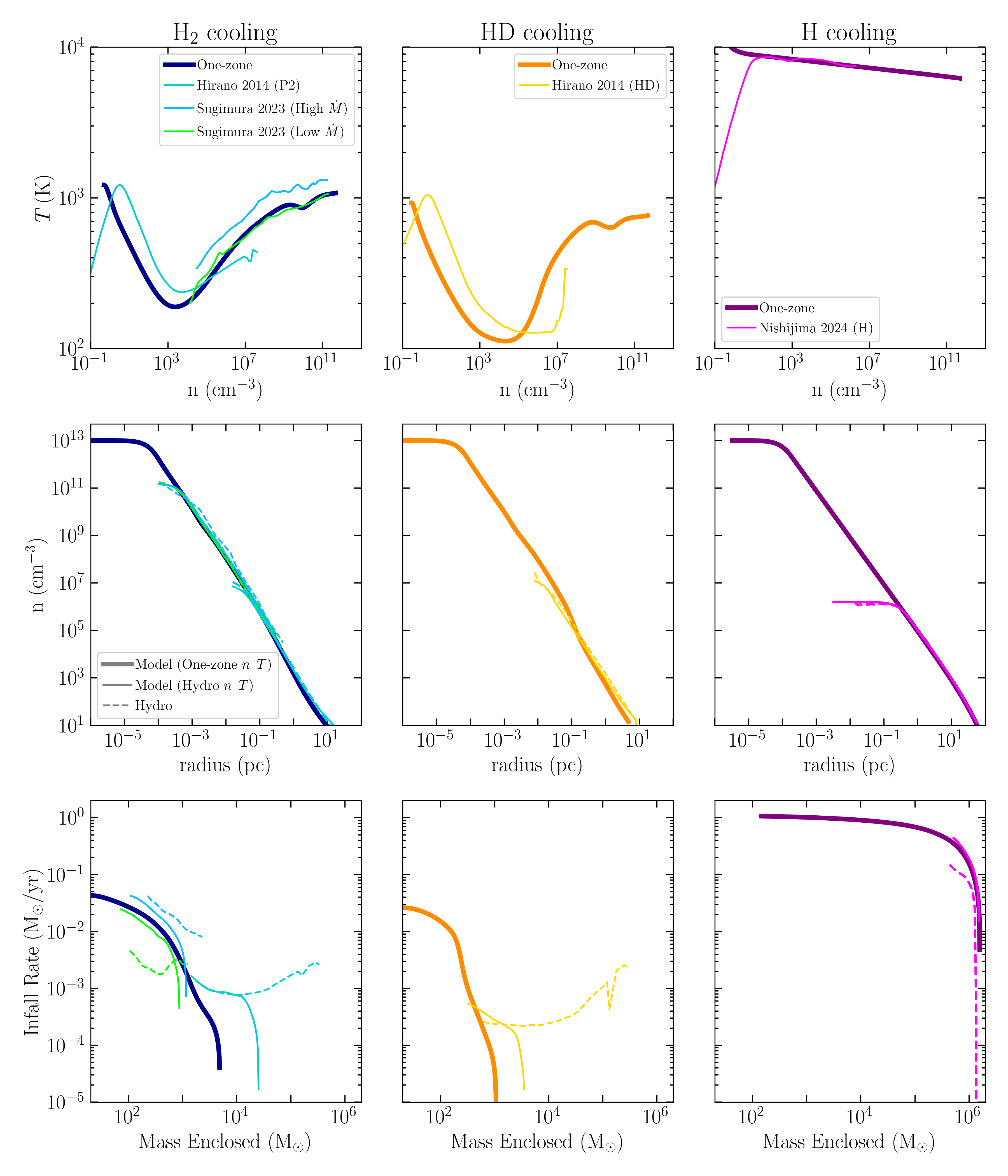}
    \caption{The density-temperature relationships (top row), density profiles (middle row) and infall rates (bottom row) for cases with $\hto$ cooling (left column), $\hd$ cooling (middle column), and $\rm H$ cooling (right column). Thick lines correspond to one-zone density temperature relationships, while thin lines indicate simulation density-temperature relationships. In the bottom two rows, solid lines indicate predictions of the model, while dashed lines indicate hydro simulation results. Because we do not model the dropoff in velocity near the core, we have truncated the simulation infall rates at $M(r= 25 r_{\rm MBE})$, with $r_{\rm  MBE}$ determined using our model.}
    \label{fig:h2pdcompare}
\end{figure*}

{In Fig.~\ref{fig:h2pdcompare} we show the thermal evolution, and associated density profiles and infall rates  both in our model (solid) and in hydrodynamics simulations (dashed) for several cases, grouped by the operative cooling mechanism. The thick lines correspond to the examples of Sections \ref{sec:eos}, \ref{sec:delay}, \ref{sec:atomic}, while thin lines correspond to simulations. For $\hto$ cooling, we consider three cases: two from the cloud-scale simulations of \cite{Sugimura2023} and one from the cosmological zoom-in simulations of \cite{hirano_one_2014}. The two cases from \cite{Sugimura2023} are snapshots when the central density is $\sim 10^{11} \, \rm cm^{-3}$ of the high infall rate and low infall rate clouds studied in that work. From \cite{hirano_one_2014} we consider the average of all clouds with intermediate infall rates (case P2 in that work) when the central density is $\sim 10^7 \, \rm cm^{-3}$. All these clouds have relatively similar thermal evolutions (Fig.~\ref{fig:h2pdcompare}, top left), and the simulated and analytic density profiles from all cases with $\hto$ cooling agree such that the different cases are difficult to visually distinguish (left middle panel). Note especially the close agreement between the one-zone model and the low infall case of \cite{Sugimura2023}, due to their similar thermal evolution. With respect to the infall rates (bottom left panel), the zero-velocity boundary condition in our model is inconsistent with the cosmological infall in \cite{hirano_one_2014}. Moreover, when the thermal evolution is taken from \cite{Sugimura2023} this boundary condition is artificially pushed to high densities/small masses because the thermal evolution is not available at low densities. Still, the model correctly determines $\sim 10^4 \, \rm M_\odot$ as the scale at which gravitational instability sets in and the large increase in infall rates between $10^{3}$ and $10^{4}\, \rm M_\odot$. The worst agreement is in the strongly rotationally supported, low-infall rate cloud from \cite{Sugimura2023}. This is reasonable, since rotation is not explicitly considered in our model (but only partially via the collapse timescale factor $f$). The observation that rotational support has a large effect on the infall velocity but a small effect on the density profile has interesting implications for future work.}

{For $\hd$ cooling (middle column of Fig.~\ref{fig:h2pdcompare}), we consider the average of all clouds with $\hd$ cooling from \cite{hirano_one_2014}, when the central density is $\sim 10^7 \, \rm cm^{-3}$. The disagreement in thermal evolution between the simulations and one-zone model propagates to the density profiles and infall rates. Still, the overall agreement and especially the distinction from the $\hto$ case is reasonable. As in the $\hto$ case, the underestimate of the temperature at low densities in the one-zone model as compared to these simulations affects the point where $\kappa_{\rm MBE}$ first exceeds unity (Rees-Ostriker criterion), which will only marginally alter the infall rate in the dense, inner region. Unfortunately, the data of \cite{hirano_one_2014} does not reveal the inner envelope structure at late times due to the low central density. \cite{hirano_one_2014} also followed the azimuthally symmetrized accretion after proto-star formation. While it is not straightforward to directly compare that calculation with our model, we point out that the proto-stellar accretion rates in that work roughly agree with the predicted infall rate at low masses in our model $\sim 10^{-2} \, \rm M_\odot/yr$ and that many of the low-mass stars (whose accretion history is less affected by feedback) exhibit the characteristic ``shoulder'' in the accretion rate predicted here by the cooling/isothermal evolution in the low density gas and heating in the high density gas (Fig.~10 in that work). }

{Finally, turning to the atomic cooling case (right column), we take a snapshot from the case of \cite{nishijima2023lowmasspopiiistar} where the Lyman-Werner background $J_{21}=30$, leading to nearly isothermal atomic cooling. In this snapshot, the central density is $\sim 10^6 \, \rm cm^{-3}$. Here, the one-zone and simulation thermal evolution agree very closely except for the initial conditions, and the density profiles are likewise extremely similar. The infall rates again agree to within a factor of few.}

{In summary, we find that our model predicts the dynamics of the collapse typically to within a factor of a few using only the thermal evolution, over some four orders of magnitude in cloud mass. The model succeeds at a level comparable to the scatter between individual simulations runs. Moreover, even a crude estimate of the thermal evolution based only on knowledge of the operative cooling mechanism only moderately degrades the accuracy, especially in the dense inner region which is most relevant for star formation. The density profiles especially are determined quite accurately in our model, supporting the notion that the modified Bonnor-Ebert mass defined in this work is indeed a critical scale in the problem. In particular, $\kappa_{\rm MBE}$ is calculated based on the density profile alone and describes the ratio between the actual density profile and the hydrostatic density profile. Thus, the mechanism by which radiative cooling establishes gravitational instability at a characteristic scale, which is a central objective of this work, is robustly demonstrated. The quantitative accuracy of the predictions can be improved both by extending the model to include effects such as rotational and turbulent support and by improving the sophistication with which the chemical-thermal evolution is estimated. These are goals for future work. }

\bsp	
\label{lastpage}
\end{document}